\documentclass[article]{aa}  

\usepackage{graphicx}
\usepackage{amsmath}
\usepackage{verbatim}
\usepackage{fancyvrb}
\usepackage{txfonts}
\usepackage{hyperref}
\usepackage{natbib}
\bibpunct{(}{)}{;}{a}{}{,} 
\usepackage{color}
\usepackage{xspace}
\usepackage{listings}
\definecolor{codegreen}{rgb}{0,0.6,0}
\definecolor{codegray}{rgb}{0.5,0.5,0.5}
\definecolor{codepurple}{rgb}{0.58,0,0.82}
\definecolor{backcolour}{rgb}{0.95,0.95,0.92}

\lstdefinestyle{mystyle}{
    backgroundcolor=\color{backcolour},   
    commentstyle=\color{codegreen},
    keywordstyle=\color{magenta},
    numberstyle=\tiny\color{codegray},
    stringstyle=\color{codepurple},
    basicstyle=\ttfamily\footnotesize,
    breakatwhitespace=false,         
    breaklines=true,                 
    captionpos=b,                    
    keepspaces=true,                 
    numbers=left,                    
    numbersep=5pt,                  
    showspaces=false,                
    showstringspaces=false,
    showtabs=false,                  
    tabsize=2
}

\newcommand{\Gaia}{\textit{Gaia}\xspace}
\providecommand\gdr[1]{\textit{Gaia}~DR#1\xspace}
\newcommand{\cat}{\ensuremath{\mathcal{C}}}
\providecommand\gbp{\ensuremath{G_\mathrm{BP}}\xspace}
\providecommand\grp{\ensuremath{G_\mathrm{RP}}\xspace}

\begin{document}

\title{Estimating the selection function of \Gaia DR3 \textit{sub}-samples}

\author{
    Alfred Castro-Ginard          \inst{\ref{inst:leiden}}\relax
\and Anthony G.A. Brown \inst{\ref{inst:leiden}}\relax
\and Zuzanna Kostrzewa-Rutkowska \inst{\ref{inst:leiden}}\relax
\and Tristan Cantat-Gaudin \inst{\ref{inst:heidelberg}}\relax
\and Ronald Drimmel \inst{\ref{inst:torino}}\relax
\and Semyeong Oh \inst{\ref{inst:cambridge}}\relax
\and Vasily Belokurov \inst{\ref{inst:cambridge}}\relax
\and Andrew R. Casey \inst{\ref{inst:monash},\ref{inst:astro3d}}\relax
\and Morgan Fouesneau \inst{\ref{inst:heidelberg}}\relax
\and Shourya Khanna \inst{\ref{inst:torino}}\relax
\and Adrian M. Price-Whelan \inst{\ref{inst:flatiron}}\relax
\and Hans-Walter Rix \inst{\ref{inst:heidelberg}}
}

\institute{Leiden Observatory, Leiden University, Niels Bohrweg 2, 2333 CA Leiden, the Netherlands\\ \email{acastro@strw.leidenuniv.nl}\relax \label{inst:leiden}
\and{Max-Planck-Institut f\"ur Astronomie, K\"onigstuhl 17, D-69117 Heidelberg, Germany \label{inst:heidelberg}}
\and{INAF - Osservatorio Astrofisico di Torino, Strada Osservatorio 20, Pino Torinese 10025 Torino, Italy \label{inst:torino}}
\and{Institute of Astronomy, University of Cambridge, Madingley Road, Cambridge CB3 0HA, United Kingdom \label{inst:cambridge}}
\and{School of Physics and Astronomy, Monash University, VIC 3800, Australia \label{inst:monash}}
\and{Centre of Excellence for Astrophysics in Three Dimensions (ASTRO-3D), Melbourne, Victoria, Australia \label{inst:astro3d}}
\and{Center for Computational Astrophysics, Flatiron Institute, 162 Fifth Ave, New York, NY 10010, USA \label{inst:flatiron}}
}

\date{Received date /
Accepted date}

\date{Received date /
Accepted date}
  
\abstract{
        Understanding which sources are present in an astronomical catalogue and which are not is crucial for the accurate interpretation of astronomical data. In particular, for the multidimensional \Gaia data, filters and cuts on different parameters or measurements introduces a selection function that may unintentionally alter scientific conclusions in subtle ways.
}{
        We aim to develop a methodology to estimate the selection function for different \textit{sub}-samples of stars in the \Gaia catalogue.
}{
       Comparing the number of stars in a given \textit{sub}-sample to those in the overall \Gaia catalogue, provides an estimate of the \textit{sub}-sample membership probability, as a function of sky position, magnitude and colour. This estimate must differentiate the stochastic absence of \textit{sub}-sample stars from selection effects.  When multiplied with the overall \Gaia catalogue selection function this provides the total selection function of the \textit{sub}-sample. 
}{
        We present the method by estimating the selection function of the sources in \Gaia DR3 with heliocentric radial velocity measurements. We also compute the selection function for the stars in the \Gaia-Sausage/Enceladus sample, confirming that the apparent asymmetry of its debris across the sky is merely caused by selection effects.
}{
        The developed method estimates the selection function of the stars present in a \textit{sub}-sample of \Gaia data, given that the \textit{sub}-sample is completely contained in the \Gaia parent catalogue (for which the selection function is known). This tool is made available in a GaiaUnlimited Python package.
}
\maketitle
%


\section{Introduction}
\label{sec:intro}

To reach meaningful scientific conclusions based on data for objects included in astronomical catalogues, we have to rely on the data and measurements these catalogues provide and, even more important, know the caveats and limitations of the catalogue. The latter aspect includes understanding what objects are not included in the catalogue, which is often characterised by the catalogue selection function $\mathcal{S}_\cat$. Selection functions are commonly constructed through either understanding of the detection efficiency and chain of procedures used to build the catalogue, or through a statistical comparison of the catalogue against a \textit{ground truth}, meaning a more complete set of sources of the same nature \citep[for a review of the basics of astronomical selection functions see][]{2021AJ....162..142R}. 

With the enormous wealth of data from recent astronomical missions, often scientific conclusions are reached based on specific \textit{sub}-samples, generated by selecting certain kinds of objects (e.g. white dwarfs, red clump stars, or stars with available velocities) based on their attributes, rather than on the full catalogue. This is often the case when working with data from the \Gaia mission \citep{2016A&A...595A...1G}, which provides astrometric and photometric measurements for  more than one billion stars in our Galaxy. In addition, it is common practice to apply additional quality cuts in order to remove undesired outliers. Every cut applied to produce a particular \textit{sub}-sample (e.g. on colour, or using data quality flags) introduces different selection effects that must be accounted for. In the case of using \Gaia data only, these selection effects can be taken into account by comparing the objects in any \textit{sub}-sample against the full \Gaia catalogue, the parent catalogue, for which the completeness and selection function is assumed to be known.

Specific efforts to estimate the selection function for \Gaia data started after the appearance of the second \Gaia data release \citep[DR2,][]{2018A&A...616A...1G}. \citet{2020MNRAS.497.1826B,2021MNRAS.501.2954B} and \citet{2020MNRAS.497.4246B} used the epoch photometry of the variable stars in \gdr{2} to estimate the \Gaia parent catalogue selection function. Building on that work, \citet{2022MNRAS.509.6205E} computed the selection function for different \textit{sub}-samples of \gdr{2} data, including the selection function of stars with heliocentric radial velocity measurements, which has also been independently estimated by \citet{2021MNRAS.500..397R}, by taking the ratios of sources with radial velocities compared to all \gdr{2} sources. To estimate the parent catalogue selection function for \gdr{3} \citep{2022arXiv220800211G}, \citet{2023A&A...669A..55C} exploited the comparison of \Gaia data with a deeper survey \citep[the Dark Energy Camera Plane Survey, DECaPS,][]{Schlafly18,2023ApJS..264...28S}, assumed to represent the \textit{ground truth} (i.e. to be 100\% complete), to estimate the completeness of \gdr{3} as a function of sky position and $G$ magnitude. This latter work, as well as the current paper, are in the context of the GaiaUnlimited project\footnote{\url{https://gaia-unlimited.org/}}, which aims at providing the community with selection functions for the different \Gaia releases, as well as for different \textit{sub}-samples of the data, together with a python package\footnote{\url{https://github.com/gaia-unlimited/gaiaunlimited}. The full documentation can be found in \url{https://gaiaunlimited.readthedocs.io/en/latest/index.html}} that contains the necessary tools for the application of different aspects of the \Gaia selection function (scanning law, \Gaia parent catalogue selection function, and several \textit{sub}-sample selection functions and how to estimate them).

\begin{figure*}[ht!]
    \includegraphics[width=\linewidth]{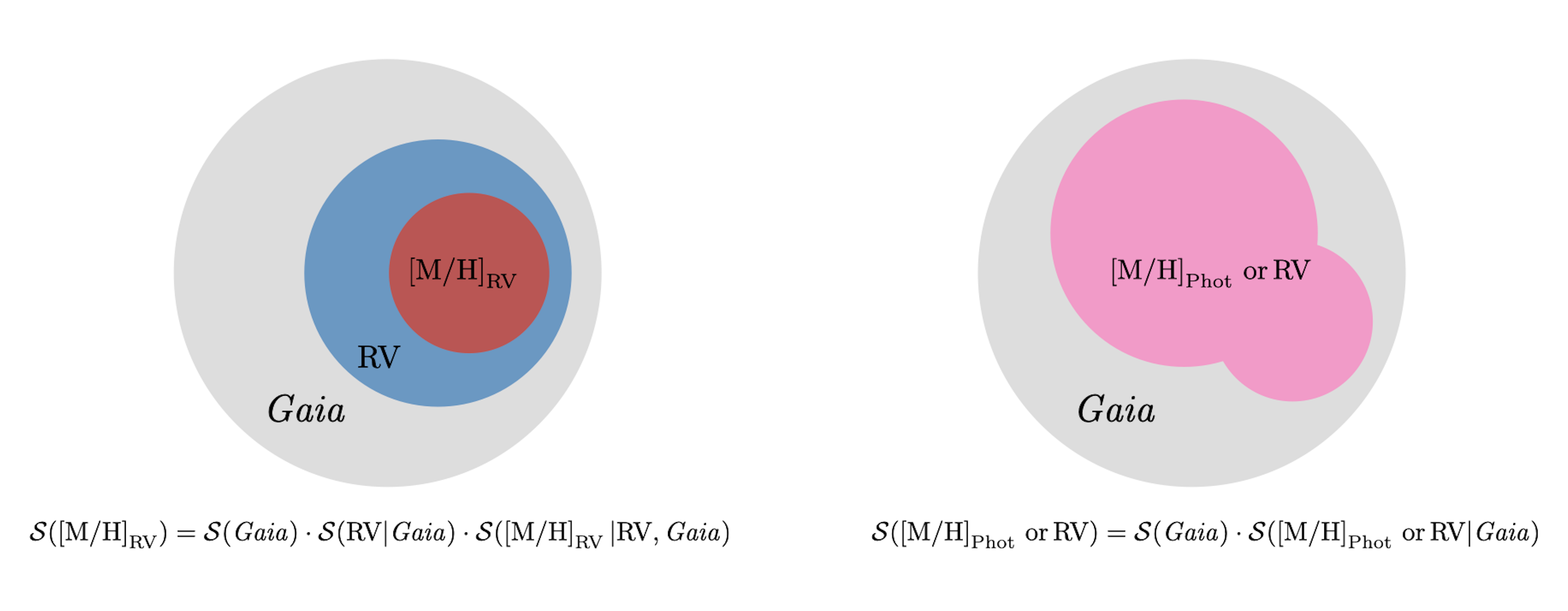}
    \caption{Sketch showing the cases of applicability of the proposed methodology.}
    \label{fig:cartoon}
\end{figure*}

The goal of this paper is to provide the means to estimate the selection function of any subset of \Gaia data. Figure~\ref{fig:cartoon} shows the cases where our methodology can be applied. The left and right panels show two examples of how to estimate the selection function when applying different filters (selection criteria) to the \Gaia catalogue, where all the sources resulting from the  filtering are included in the parent catalogue. In both cases, all the subsets shown can be drawn from simple queries to the \Gaia archive. We stress again that we require the \textit{sub}-sample to be entirely contained in the \Gaia source catalogue, for which the selection function was empirically modelled by \citet{2023A&A...669A..55C}.

This paper is organised as follows. In Sect.~\ref{sec:method}, we describe the methodology used to estimate the selection function of a subset of \Gaia data. We apply the method to the stars with heliocentric radial velocity measurements in \Gaia DR3 and compare our results with previous similar methods in Sect.~\ref{sec:rvs}. Section~\ref{sec:GES} shows how to use the estimated selection function in a real science case, the \Gaia-Sausage/Enceladus sample. Finally, in Sect.~\ref{sec:conclusions} we discuss our conclusions. We also provide examples of the queries made in the \Gaia archive in Appendix~\ref{app:appendix}, an example of python code to generate \textit{sub}-sample selection functions using the GaiaUnlimited python package in Appendix~\ref{app:github} and the selection function for different relevant subsets in Appendix~\ref{app:examples}.

\section{Method}
\label{sec:method}

Here we present a method to estimate the selection function of \Gaia catalogue \emph{sub}-samples. These can be subsets drawn directly from the \Gaia catalogue, or a set of sources included in another survey that was exclusively selected from the \Gaia catalogue (for example, a spectroscopic survey that draws its targets from \Gaia). Our method relies on the fact that the \Gaia catalogue is the parent catalogue to these \textit{sub}-samples, and that its basic selection function has already been well-characterised. Generalising the cases sketched in Fig.~\ref{fig:cartoon}, the probability $\mathcal{S}_\cat(\mathbf{q})$ that a source makes it into our \textit{sub}-sample is described by \citep[see section 2.1 and equation 2 in][]{2021AJ....162..142R}
\begin{equation}
\label{eqn:SF}
\mathcal{S}^\mathrm{subsample}_\cat(\mathbf{q}) = 
\mathcal{S}_\cat(\mathbf{q} \, | \, \mathbf{q} \text{ in parent}) \,
\cdot \,
\mathcal{S}^\mathrm{parent}_\cat(\mathbf{q}),
\end{equation}
where $\mathcal{S}^{\text{parent}}_\cat(\mathbf{q})$ describes the probability that a source with attributes $\mathbf{q} = \{\ell,b,G, \ldots\}$ will make it into the \Gaia catalogue and $\mathcal{S}_\cat(\mathbf{q} \, | \, \mathbf{q} \text{ in parent})$ is the probability that a source will be in the \textit{sub}-sample, given that it is in the \Gaia parent catalogue. The method we developed focuses on the estimation of \mbox{$\mathcal{S}_\cat(\mathbf{q} \, | \, \mathbf{q} \text{ in parent})$}, which then becomes a multiplicative factor to the parent catalogue selection function provided by \citet{2023A&A...669A..55C} in estimating the total selection function of our \textit{sub}-sample.

\subsection*{Probability of selecting the sources in the \textit{sub}-sample}
\label{sec:binomial}

We model the number of sources that end up in our \textit{sub}-sample as a Binomial distribution, which assumes that sources are randomly selected with a given probability which depends on the source attributes $\mathbf{q}$. The Binomial distribution is given by
\begin{gather}
\label{eqn:binomial}
Y \sim \text{Binomial}(n,p)\,,\nonumber\\
P(Y = k) = {n\choose k}\,p^k\,(1-p)^{n-k}\,,
\end{gather}
where $n$ is the number of sources in the \Gaia catalogue with attributes $\mathbf{q}$, $k$ is the number of sources with the same attributes that are contained in our \textit{sub}-sample and $p$ is the probability that a source makes it into our \textit{sub}-sample. 

We estimate the value of $p$ from the known values of $n$ and $k$ using a Bayesian approach. To estimate the posterior probability of $p$, we choose as a prior the Beta distribution. This is a common choice since it is a conjugate prior probability distribution for the Binomial distribution, meaning that the posterior probability of $p$ is also a Beta distribution which is updated according to the data. We use an uninformative uniform prior distribution, which means a Beta($\alpha$,$\, \beta$) distribution function with $\alpha = \beta = 1$. In this particular case and considering the above assumptions, the posterior distribution of $p$ is given by Beta($k+1,n-k+1$) which has a mean value of
\begin{equation}
\label{eqn:mean}
\text{E}(p) = \frac{k+1}{n+2}
\end{equation}
and tends to $k/n$ as $k$ and $n$ get larger. The variance of the Beta($k+1,n-k+1$) distribution function is given by
\begin{equation}
\label{eqn:variance}
\text{var}(p) = \frac{(k+1)(n-k+1)}{(n+2)^2(n+3)}.
\end{equation}
We have summarized the full posterior distribution function in the Eqn.~\ref{eqn:mean} and \ref{eqn:variance}. However, the advantage of using Bayesian statistics is that we have access to the full posterior distribution function for the probability $p$ which, as already mentioned, is given by Beta($k+1,n-k+1$) in this case.

To apply the above method the parent catalogue and \textit{sub}-sample data are both binned by the attributes $\mathbf{q}$ and $n$ and $k$ are recorded for each bin, from which $p$ and its variance are then estimated according to the equations above. We then take $\mathcal{S}_\cat(\mathbf{q} \, | \, \mathbf{q} \text{ in parent}) = \text{E}(p)$.  We note here that the parent catalogue selection function may explicitly depend on only a subset $\mathbf{q}'$ of the attributes $\mathbf{q}$ used to select the \textit{sub}-sample. It is assumed that $\mathcal{S}_\mathcal{C}^\mathrm{parent}(\mathbf{q})=\mathcal{S}_\mathcal{C}^\mathrm{parent}(\mathbf{q}')$. This is illustrated in the next section.

\begin{figure}[ht!]
\centering
\includegraphics[width = 1.\columnwidth]{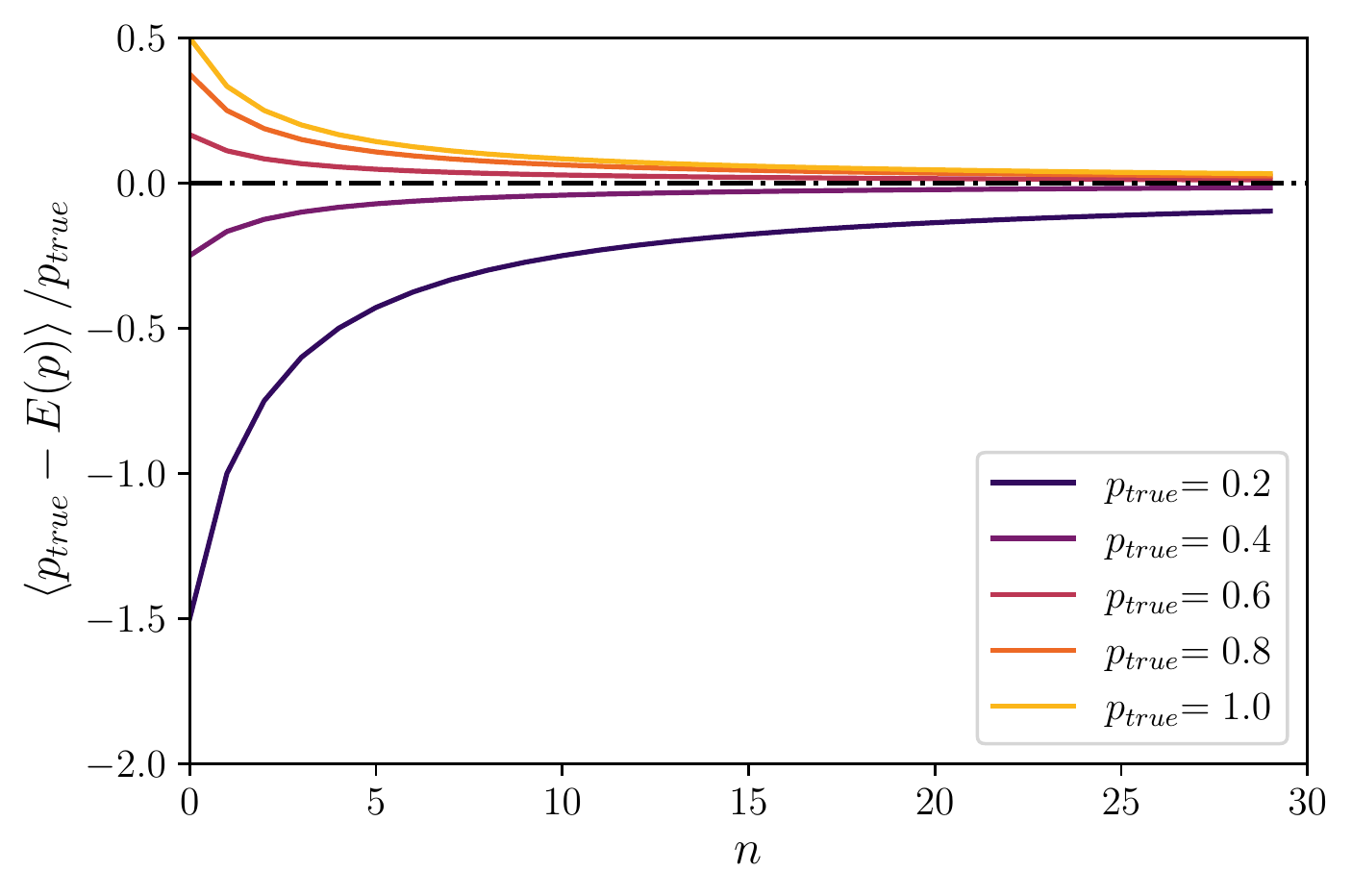}
\caption{Bias of the probability estimator described by Eq.~\ref{eqn:mean} for low values of $n$. The dash-dotted black line corresponds to an unbiased estimator, and the solid lines with different colours represent the experiment for different true probabilities $p_\text{true} \geq 0.5$.} 
\label{fig:bias}
\end{figure}

In the limit of many stars, this estimate simply becomes the ratio of \textit{sub}-sample-to-total \Gaia stars. But if the number of \textit{sub}-sample stars is small (or even zero) we must differentiate whether this is because of selection effects or simply reflects the stochasticity of the sampling. Indeed, estimating the selection probability from the expected value given by Eq.~\ref{eqn:mean} may produce biased results, particularly when both $k$ and $n$ are small. This is captured in the variance of the posterior distribution described in Eq.~\ref{eqn:variance} (for low values of $k$ and $n$, the variance will be higher and therefore the selection probability $p$ is less constrained). To provide better insight into this, we evaluate in Fig.~\ref{fig:bias} the bias of our estimator for different {\em true} probabilities $p$ as a function of $n$. As expected, the bias of our estimate increases as $n$ decreases and tend to zero for high values of $n$. Figure~\ref{fig:bias} can help us fix a minimum value of stars in the \Gaia catalogue per bin ($n$ in our notation). For instance, for $n \sim 20$ stars, the maximum bias expected is around $5\%$. In the case of $p_\text{true} < 0.5$, the expected E$(p)$ can be severely overestimated for small $n$ and, therefore, bins containing larger values of $n$ must be used.  The suitable choice of bins to avoid these biases in the selection function estimate must be vetted for each application.

\section{The selection function for stars with a heliocentric radial velocity in \Gaia DR3}
\label{sec:rvs}

We now apply the method described in Sect.~\ref{sec:method} to the sample of \gdr{3} sources with available heliocentric radial velocity measurements \citep{2022arXiv220605902K}. To generate the data for estimating this selection function, we query the \gdr{3} archive for the number of stars with heliocentric radial velocity (RV) measurements as well as the number of stars in the \gdr{3} parent catalogue ($k$ and $n$ in our notation, respectively). In Appendix~\ref{app:appendix} we included an example of the query to retrieve the RV \textit{sub}-sample in the desired format and an example of the resulting output is shown in Table~\ref{tab:resultquery}. We bin the data according to sky position (HEALPix), magnitude and colour bins, and provide the selection function in every bin where both $k$ and $n$ are available. In the context of GaiaUnlimited, the RV selection function is provided in the Python package as \texttt{DR3RVSSelectionFunction}, corresponding to precomputed sky maps at the resolution of HEALPix level $5$, in $0.2$ mag wide bins in $G$ and $0.4$ mag in $G-\grp$. As noted above, we assume here that $\mathcal{S}_\mathcal{C}^\mathrm{parent}(\ell,b,G,\grp)=\mathcal{S}_\mathcal{C}^\mathrm{parent}(\ell,b,G)$. Nevertheless, the selection function of the RV sample will be strongly dependent on the colour. The explicit $G - \grp$ dependence of the RV selection function is because the publication of RV measurements depends on the sources having an estimation of their $G_{RVS}$ magnitude and their effective temperature \citep{2022arXiv220605725S}. Both requirements can be well captured using as a proxy the $G - \grp$ colour. Also, using $G - \grp$ instead of $\gbp - \grp$ is preferred due to the known callibration issues of $\gbp$ at the faint end \citep{2021A&A...649A...3R}. 

\begin{figure*}[ht!]
\centering
\includegraphics[width = 1.\textwidth]{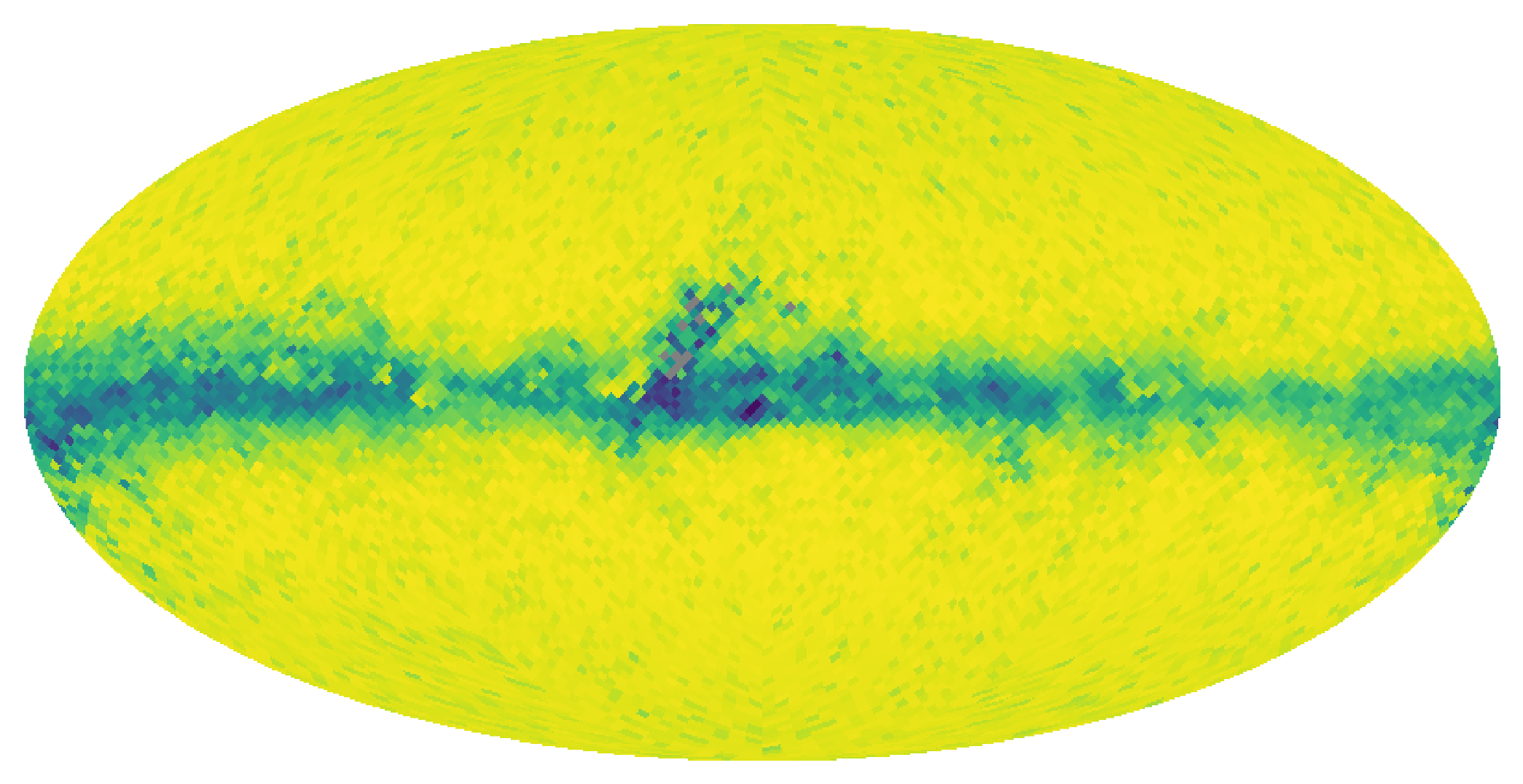}
\includegraphics[width = 1.\textwidth]{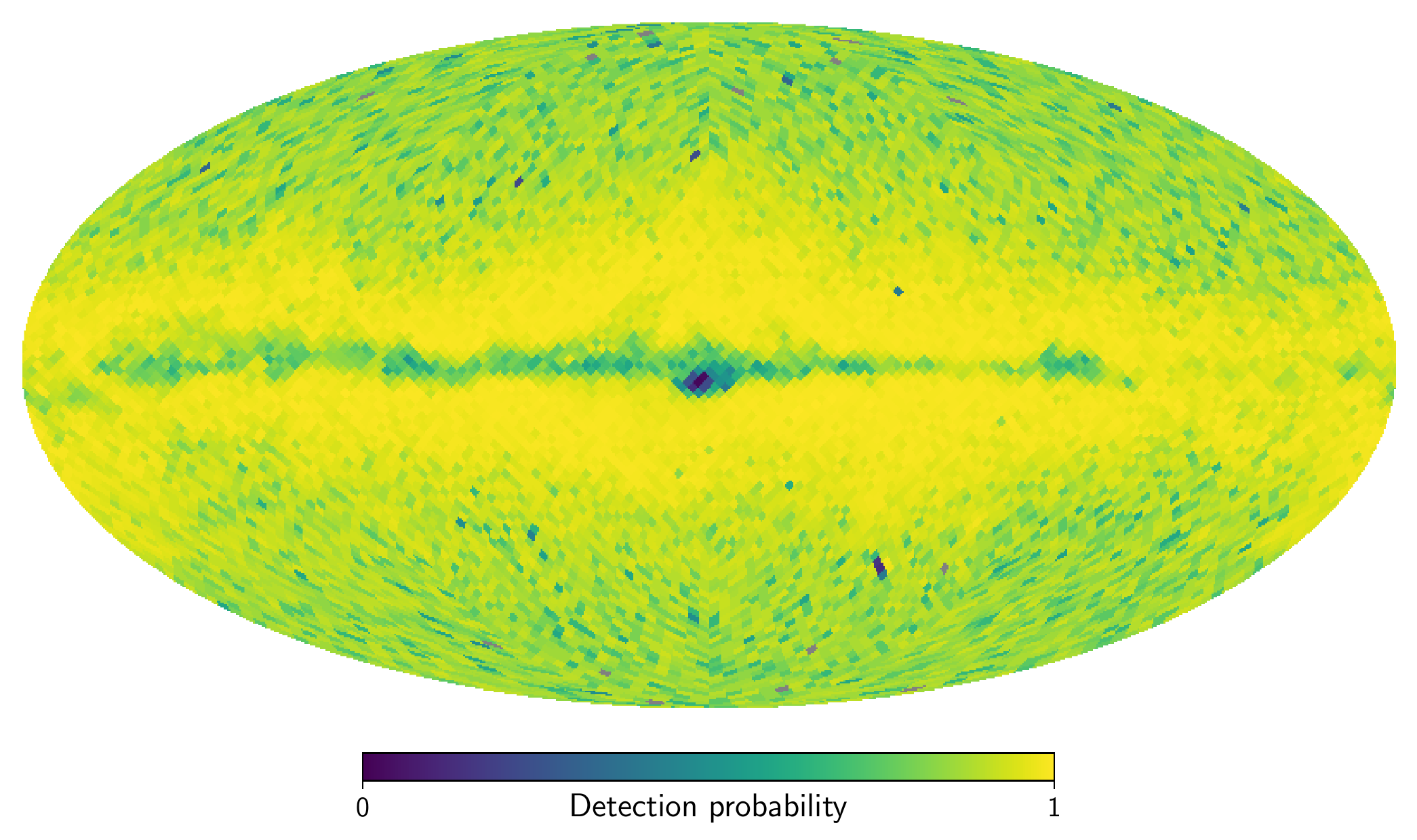}
\caption{Sky maps of the selection function  for sources with available radial velocities at magnitude $G = 13$ and colour $G-\grp = 0.5$ (top panel), and $G = 14$ and colour $G-\grp = 1$ (bottom panel). These maps are shown at HEALPix level $5$, with $0.2$ mag bins in $G$ and $0.4$ mag bins in $G-\grp$. They    manifestly depend on both magnitude and colour.}
\label{fig:rvs}
\end{figure*}

\begin{figure}[ht!]
\centering
\includegraphics[width = 1.\columnwidth]{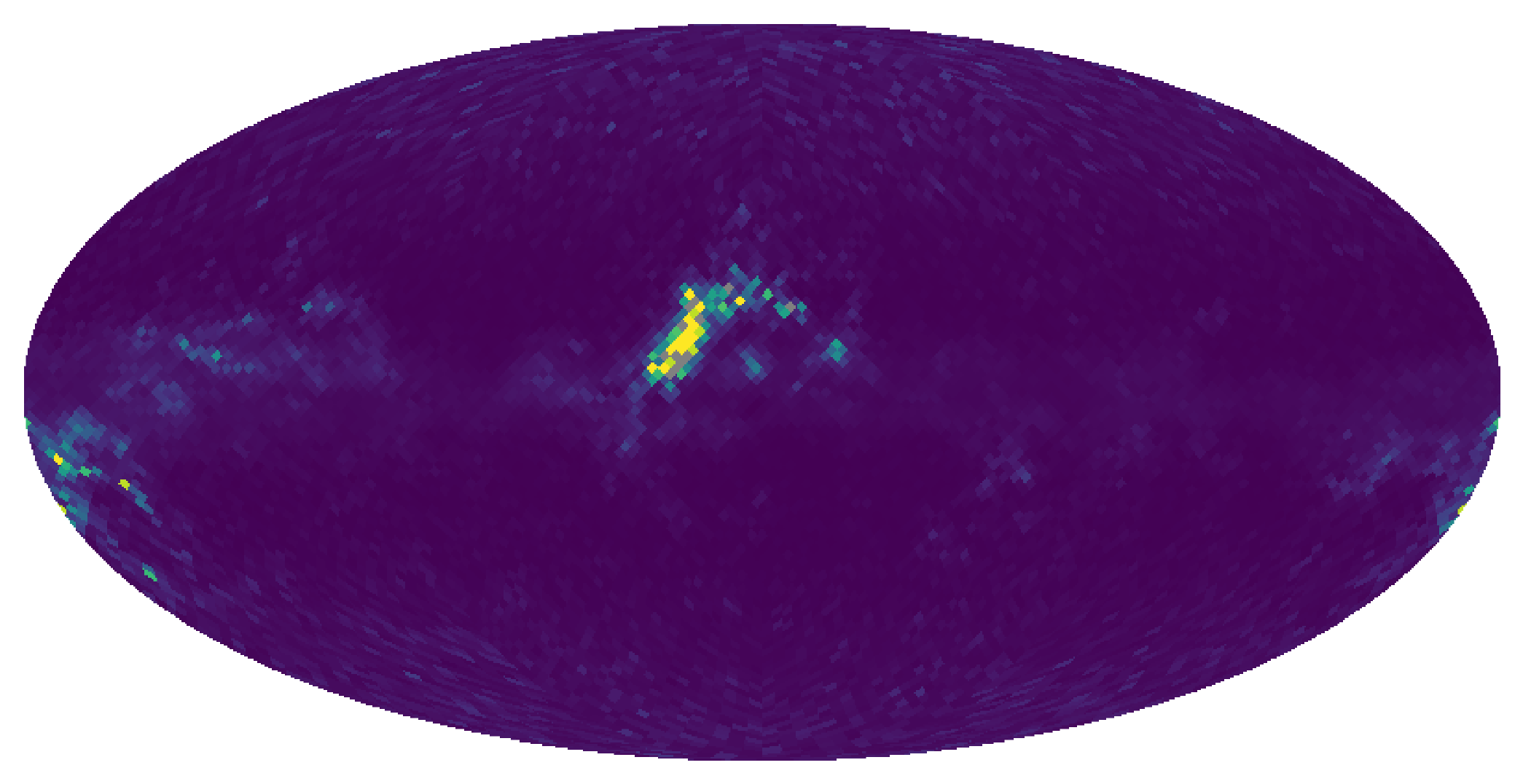}
\includegraphics[width = 1.\columnwidth]{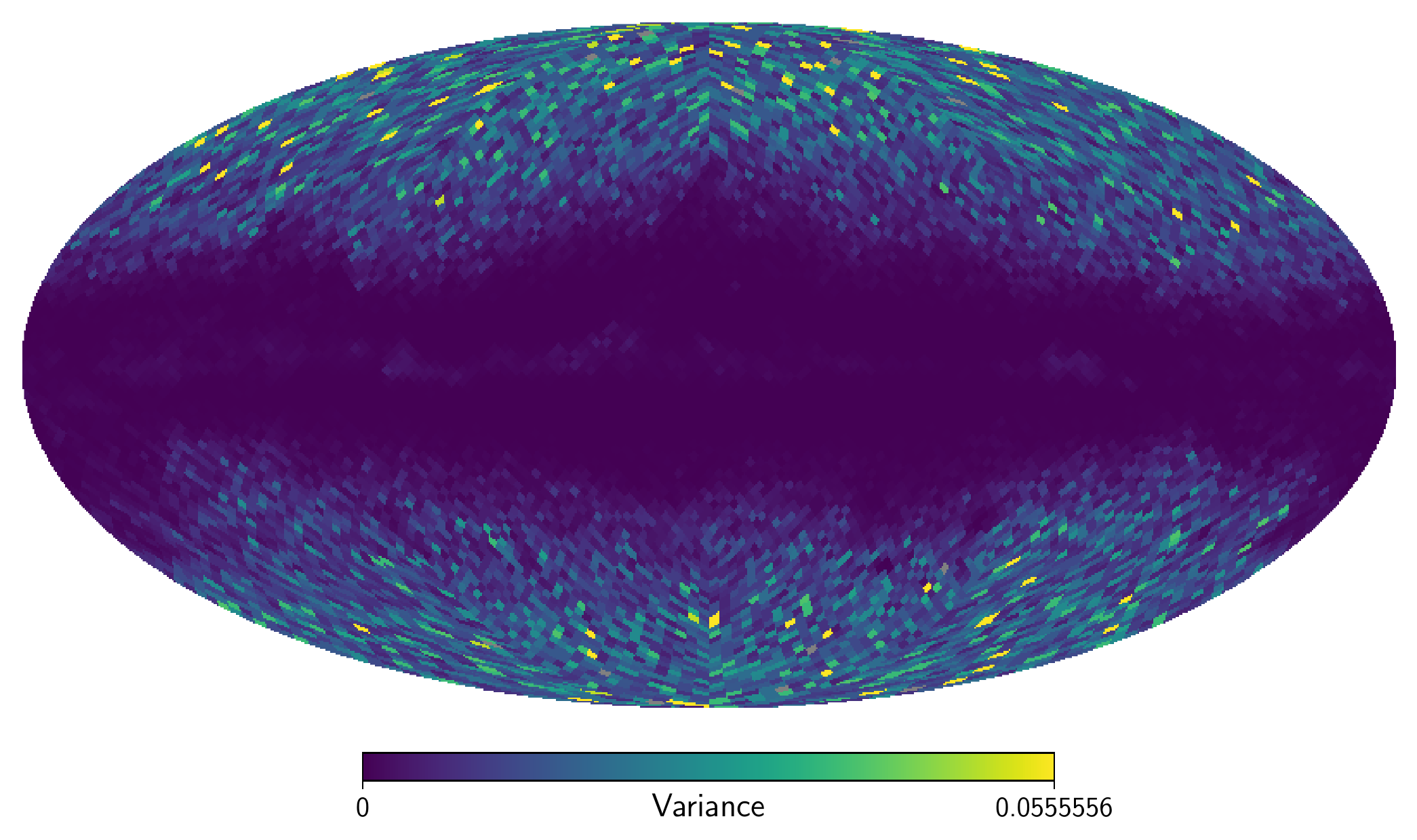}
\caption{Uncertainty in the selection function of stars with RV given by the variance of the posterior probability distribution function. The top panel corresponds to $G = 13$ and colour $G-\grp = 0.5$, and the bottom panel to $G = 14$ and colour $G-\grp = 1$. The sky maps correspond to the resolution of HEALPix level $5$.}
\label{fig:rvs_var}
\end{figure}

Figure~\ref{fig:rvs} shows sky maps of the RV selection function at magnitude $G = 13$ and $G-\grp = 0.5$ in the top panel and $G = 14$ and $G-\grp = 1$ in the bottom panel, calculated according to Eq.~\eqref{eqn:SF}. Note that in this case the term $\mathcal{S}^{\text{parent}}_\cat(\mathbf{q})$ describing the parent catalogue selection function is always $1$ (in both cases) due to the bright $G$ magnitude limit of the RV \textit{sub}-sample\footnote{In other words: in the regions of the parameter space where \Gaia radial velocities are available, the parent \Gaia~DR3 catalogue is complete.}. We find low selection probability in the Galactic midplane, particularly in the Galactic centre where the crowding effects are important. In the case of $G = 14$ and $G-\grp = 1$, the selection probability decreases as Galactic latitude increases, the selection function in these regions is underestimated due to noisy estimations of the selection function given the small values of both $n$ and $k$ for these extreme values of $G$ and $G-\grp$ (see Fig.~\ref{fig:bias} for an estimation of the bias as a function of $n$). We show the statistical uncertainty on both estimates in Fig.~\ref{fig:rvs_var}. A large number of sources near the Galactic plane makes the uncertainty (computed as the variance of the posterior probability distribution function in each HEALPix region) significantly smaller than at high Galactic latitudes, except for highly obscured regions.

\begin{figure}
    \centering
    \includegraphics[width = 1.\columnwidth]{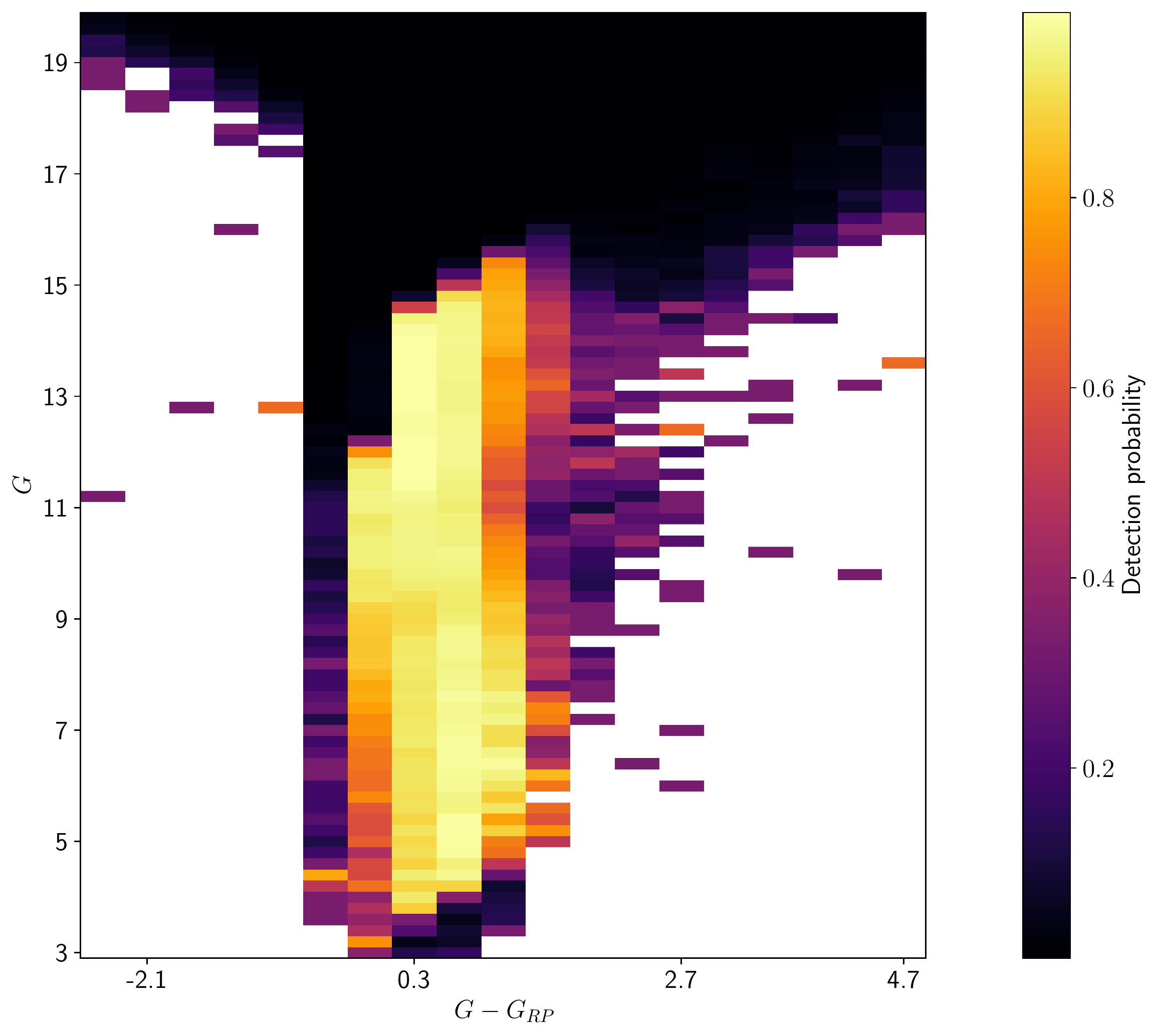}
    \caption{Detection probability for the sources with available RV measurements at latitudes $|b| > 30\deg$, as a function of $G$ magnitude and $G-\grp$ colour. The width of the bins is $0.2$ mag in $G$ and $0.4$ mag in $G-\grp$.}
    \label{fig:RV_g_grp}
\end{figure}

To avoid bias and large uncertainties in an empirically evaluated completeness map it is necessary to assure that a sufficient number of sources are in the bins used to evaluate the selection function. In Fig.~\ref{fig:rvs} we have binned in magnitude, colour and by HEALPix. One strategy to mitigate the problem of small number statistics, for example, is to adopt a sky map with variable resolution, adopting larger areas at high latitudes where there are fewer stars. Indeed, because of the bright magnitudes of the RV sample, the only dependence of the selection function on direction is due to crowding, and the selection function at high latitudes is only a function of $G$ and $G-\grp$.  In Fig.~\ref{fig:RV_g_grp} we show the selection function for the RV sample as a function of $G$ and $G-\grp$ for $b > 30\deg$ and $b < 30\deg$.

The selection function of the RV sample has dramatically improved in \Gaia DR3 compared to EDR3 (where the radial velocities were inherited from \Gaia DR2). The magnitude limit in this sample has increased from $G_{\text{RVS}} = 12$ mag in \Gaia EDR3 to $G_{\text{RVS}} = 14$ mag in \Gaia DR3, resulting in a total of $\sim 33$ million sources in the last data release compared to the $\sim 7$ million in EDR3. In Appendix~\ref{app:dr2_vs_dr3}, we show a comparison of the improvement of the RV sample in \Gaia DR3 with respect to \Gaia EDR3, at magnitude $G = 13$ mag.

\subsubsection*{Comparison to \citet{2022MNRAS.509.6205E}}
\label{sec:comparison}

\citet{2022MNRAS.509.6205E} estimated the selection function for three specific subsets of the \Gaia EDR3 release. The authors provide, through precomputed sky maps, the probability that a source contained in \Gaia EDR3, has i) a reported parallax and proper motion, ii) RUWE below $1.4$ and iii) a reported RV measurement (from \Gaia DR2) as a function of sky position, $G$ magnitude and $G-\grp$ colour (with this last dependence only for ii and iii). In short, their methodology describes the subset selection function as a sum of needlets across the sky where their coefficients are modelled by a Gaussian process prior in magnitude and colour \citep[see][for a detailed description]{2022MNRAS.510.4626B}. The use of needlets introduces spatial smoothing instead of estimating individual independent probabilities in each bin, which avoids being dominated by noisy data. Similarly, the Gaussian processes introduce a correlation in the magnitude and colour dimensions. 

As in our method described in Sect.~\ref{sec:binomial}, the core assumption of \citet{2022MNRAS.509.6205E} is that the probability to sample $k$ stars out of $n$ (from the parent catalogue) is described by the Binomial likelihood distribution with a Beta uniform distribution prior. Both approaches use the same data as the starting point (see Appendix~\ref{app:appendix}).
Compared to our ratio-based method, the complex statistical model developed by \citet{2022MNRAS.509.6205E} comes with the advantage of providing an estimate of the selection function even when no data is available in a certain bin, and a more robust estimation for bins with a low number of stars. However, their forward-modelling approach is significantly more computationally expensive. While our running time is defined by the time of the query to the \Gaia archive  (typically of the order of tens of minutes), the statistical model in \citet{2022MNRAS.509.6205E} runs for approximately one week when parallelised over $88$ cores, making the computation of custom \textit{sub}-sample selection functions impractical. 

\begin{figure*}[ht!]
\centering
\includegraphics[width = 1.\textwidth]{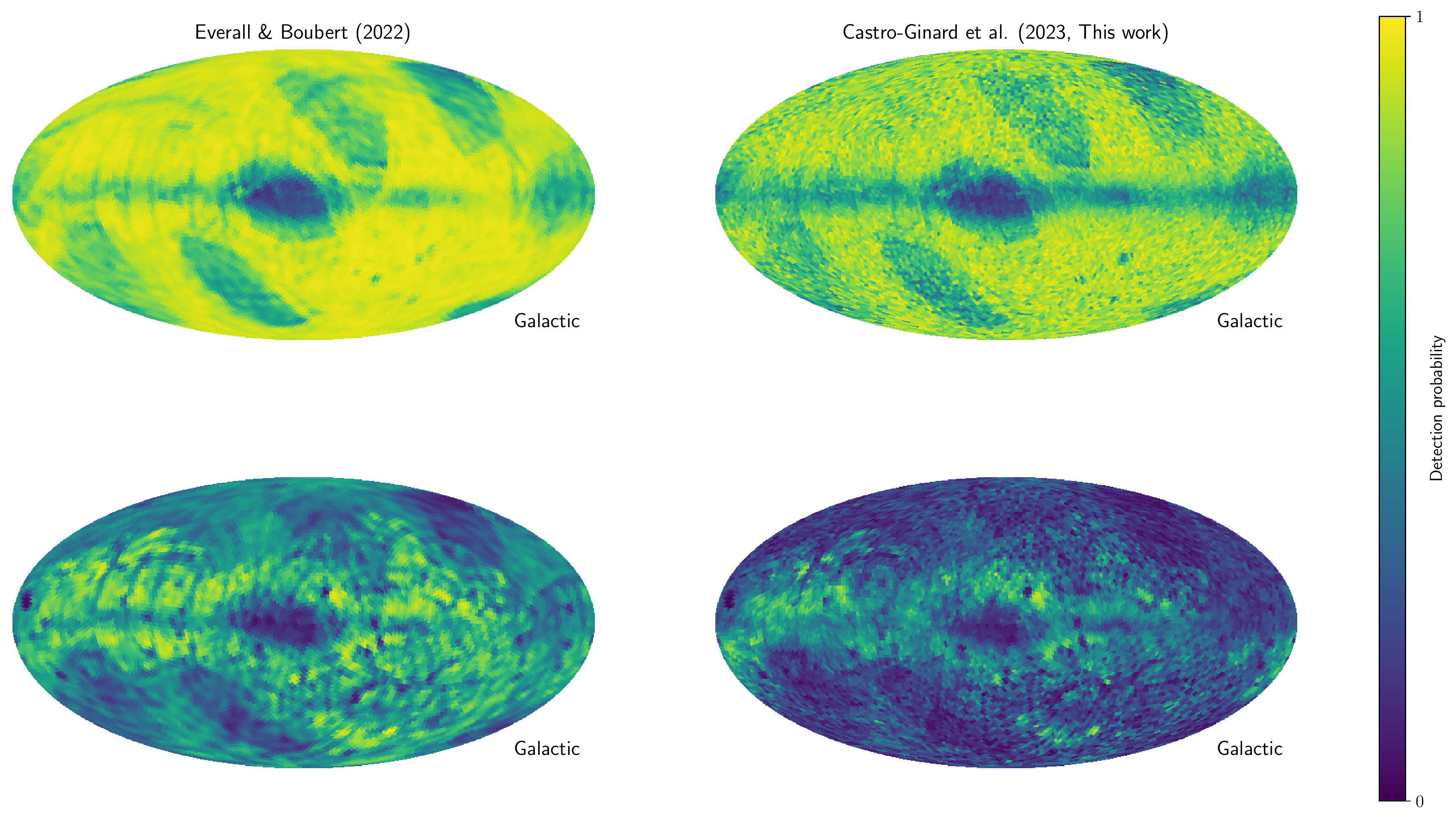}
\caption{Comparison of the method by \citet[][left column]{2022MNRAS.509.6205E} and the one developed in this paper (right column). The top panels show the \Gaia EDR3 selection function for stars with radial velocities at magnitude $G = 12$ mag, while the bottom panels show the same selection function at magnitude $G = 13$ mag. Both methodologies show a good general agreement in the results, capturing similar features in the sky maps, with the method developed by \citet{2022MNRAS.509.6205E} being a smooth version of the maps due to the inclusion of smoothing and correlation coefficients between different bins. All the maps correspond to a resolution of HEALPix level $5$.}
\label{fig:compare_method}
\end{figure*}

In order to compare the method described by \citet{2022MNRAS.509.6205E} and the method we developed, we estimate the completeness of the sources with radial velocities in \Gaia EDR3 (inherited from \Gaia DR2) using both methodologies (the actual condition is \texttt{dr2\_rv\_nb\_transits >= 4}), which should provide similar results. Figure~\ref{fig:compare_method} shows sky maps of the selection function estimated with \citet{2022MNRAS.509.6205E} method (left columns) and ours (right columns). We see a general agreement in the main features, meaning the imprint of the scanning law for $G = 12$ mag or the Initial \Gaia Source List (IGSL) at the faint end ($G = 13$ mag), top and bottom rows respectively, with our method being noisier due to the lack of smoothing between different bins. Given the similarities in both methodologies, we confirm that they provide similar results with our method being a fast alternative to compute the selection function for any \textit{sub}-sample of \Gaia data. However, there is a notable offset which might in part be explained by the bias in our estimator (see section \ref{sec:binomial}).

\section{Selection function for Gaia-Enceladus}
\label{sec:GES}

\begin{figure}[ht!]
\centering
\includegraphics[width = 1.\columnwidth]{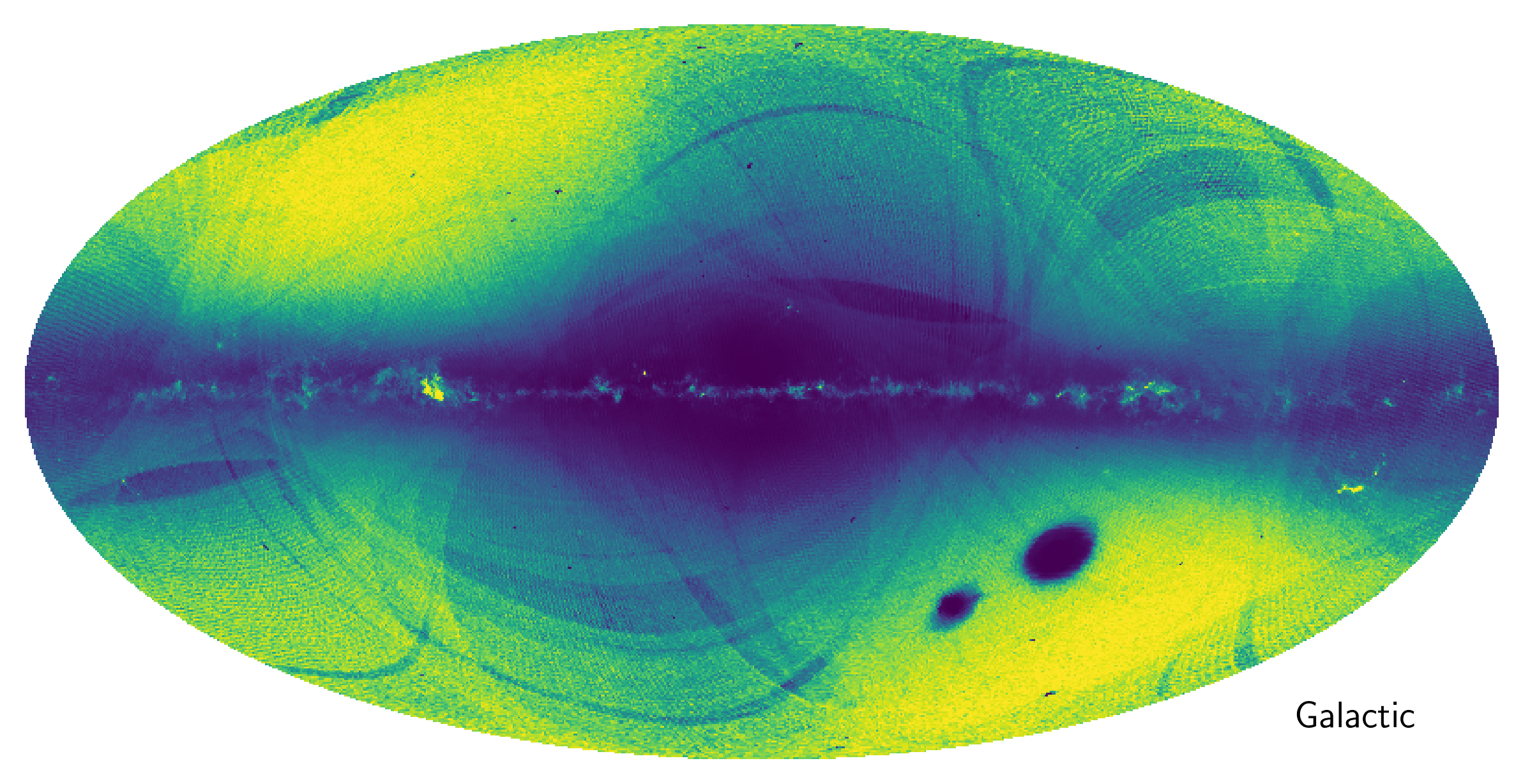}
\includegraphics[width = 1\columnwidth]{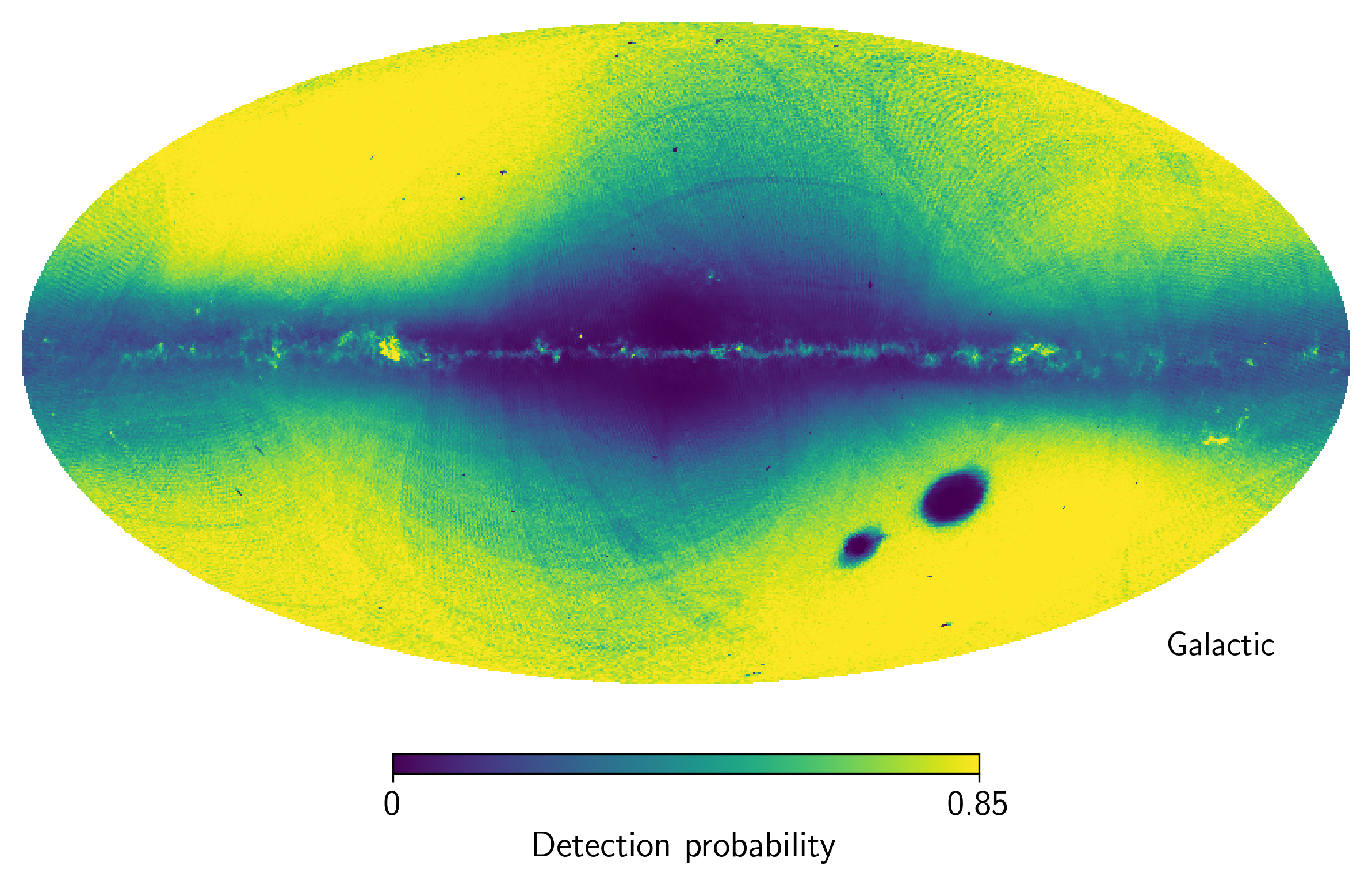}
\caption{Selection function for sources with $\varpi > 0.1$ mas and $\varpi/\sigma_\varpi > 5$ mas. The top panel shows the case for \Gaia DR2, while the bottom panel shows \Gaia DR3. Both maps correspond to the resolution of HEALPix level $7$. The magnitude and colour dependencies have been marginalised out.}
\label{fig:SF_GES}
\end{figure}

With the advent of \gdr{2}, and using the $\sim 7$ million sources with radial velocities, \citet{2018Natur.563...85H} reported a retrograde kinematic stellar structure in the nearby halo, dubbed \Gaia-Sausage/Enceladus (GS/E), which traces a major accretion event experienced by the Milky Way that contributed to the formation of its thick disc \citep[see][for details on its discovery in \Gaia DR1]{2018MNRAS.478..611B}. \citet{2018Natur.563...85H} selected stars belonging to GS/E as a set of cuts in the \Gaia DR2 catalogue with available radial velocities to show the structure of its debris. These cuts include $\varpi > 0.1$ mas, $\varpi/\sigma_{\varpi} > 5$, and $-1500 < \text{L}_{\text{z}} < 150$ kpc km$/$s. The authors found that the GS/E debris covers the whole sky, with an asymmetric shape for the more distant stars ($0.1 < \varpi < 0.25$ mas, see their Fig.~3). Each of these cuts introduces a selection effect that can be accounted for when computing the selection function. As pointed out by \citet{2018Natur.563...85H}, we find that the main source of the observed asymmetry in the GS/E debris is the selection effect caused by both the cuts in $\varpi$ and $\varpi/\sigma_{\varpi}$. We used the method described in Sect.~\ref{sec:method} to estimate the selection function of the stars in both \Gaia DR2 and DR3 that satisfy the two parallax cuts. The result is shown in Fig.~\ref{fig:SF_GES} as a function of sky position (the magnitude and colour dependencies have been marginalised out). The asymmetry (from top-left to bottom-right) is visible for both the \Gaia DR2 and DR3 \textit{sub}-samples, although much less prominent in DR3. The imprint of the scanning law is less pronounced for DR3 as well, as expected due to its longer observational baseline and more homogeneous coverage of the celestial sphere.

\begin{figure}[ht!]
\centering
\includegraphics[width = 1.\columnwidth]{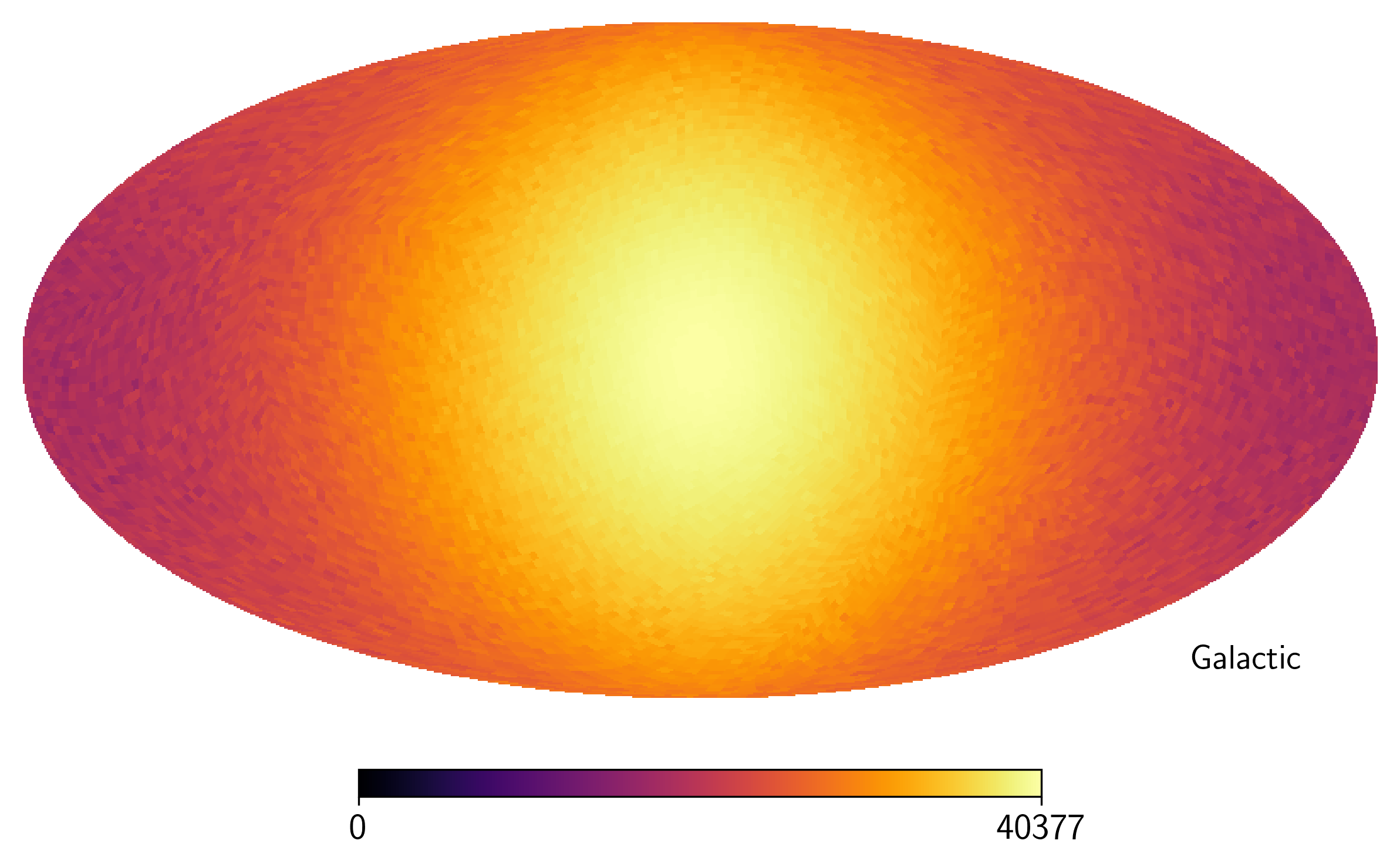}
\includegraphics[width = 1\columnwidth]{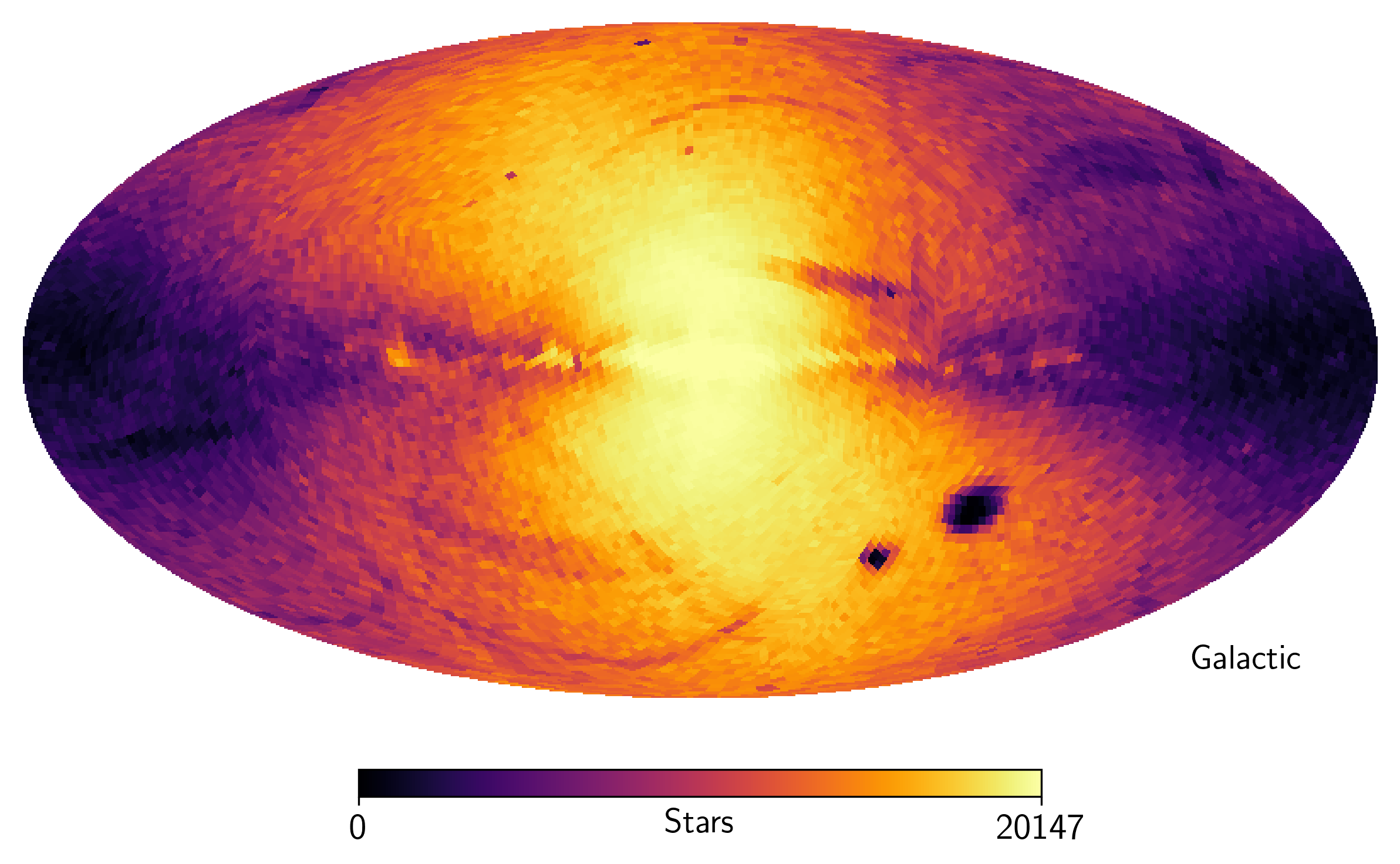}
\caption{Simulated sample of RGB stars before (top panel) and after (bottom panel) the application of the selection function accounting for the $\varpi$ and $\varpi/\sigma_\varpi$ selection effects. The asymmetry on the sample after the application of the selection function can be seen from the top-left to the bottom-right of the plot. The maps are computed at HEALPix level $5$.}
\label{fig:simu_SF}
\end{figure}

In order to check if the selection effects from the cuts in $\varpi$ and $\varpi/\sigma_\varpi$ can account for the asymmetry seen in the GES sample, we simulated a spherical distribution of red giant branch stars (RGBs) in the halo. The simulated distribution is shown in the top panel of Fig.~\ref{fig:simu_SF}. We then apply the selection function represented in Fig.~\ref{fig:SF_GES} (corresponding to \Gaia DR2) at different magnitude bins, and estimate the total expected number of stars in the corrected sample. This is shown in the bottom panel of Fig.~\ref{fig:simu_SF}, where we can see that the application of the selection effects in these two dimensions resulted in an asymmetric distribution of the initially spherical distribution of RGB stars, confirming that this asymmetry is merely a selection effect.

\begin{figure}
\centering
\includegraphics[width = 1\columnwidth]{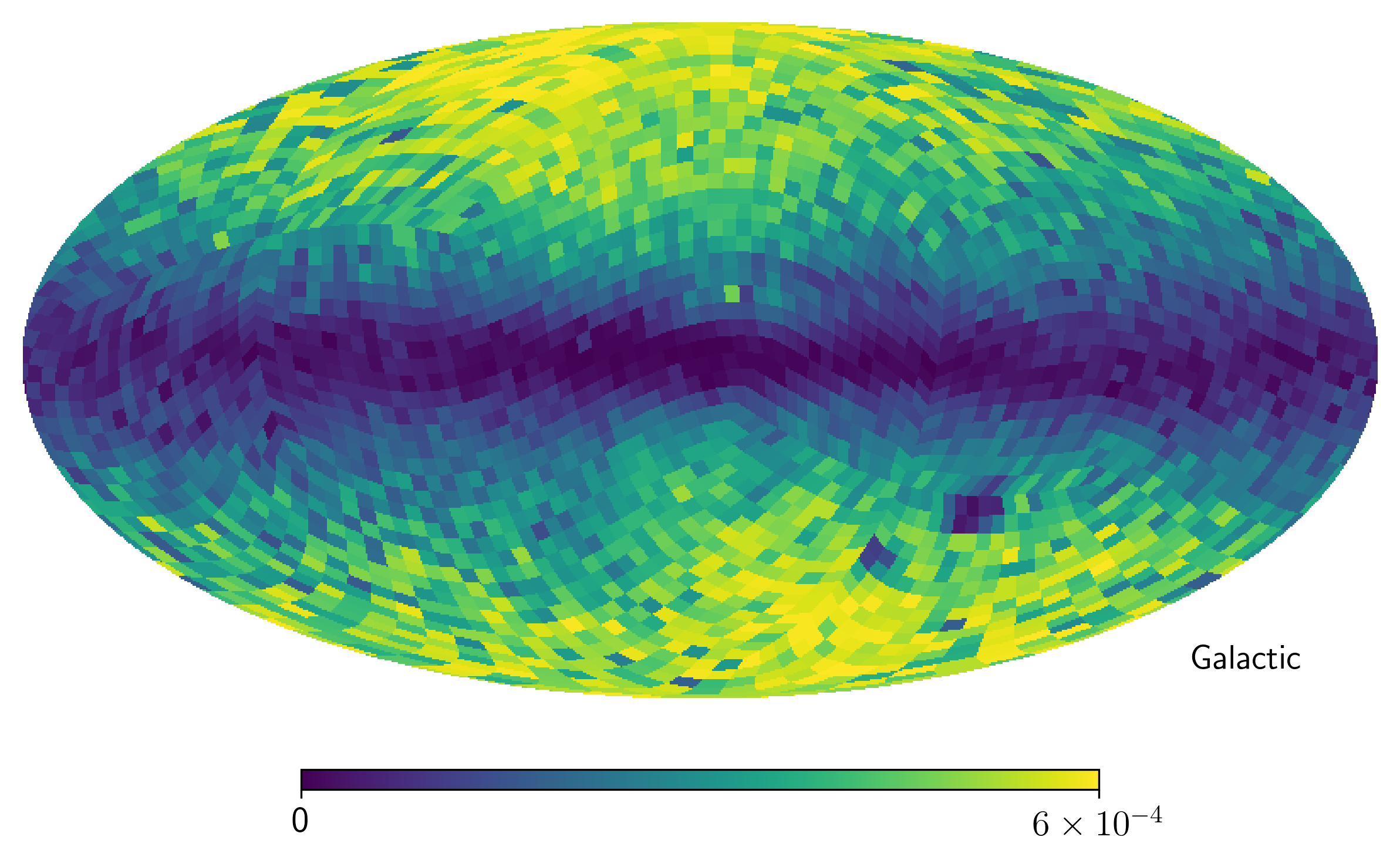}
\includegraphics[width = 1\columnwidth]{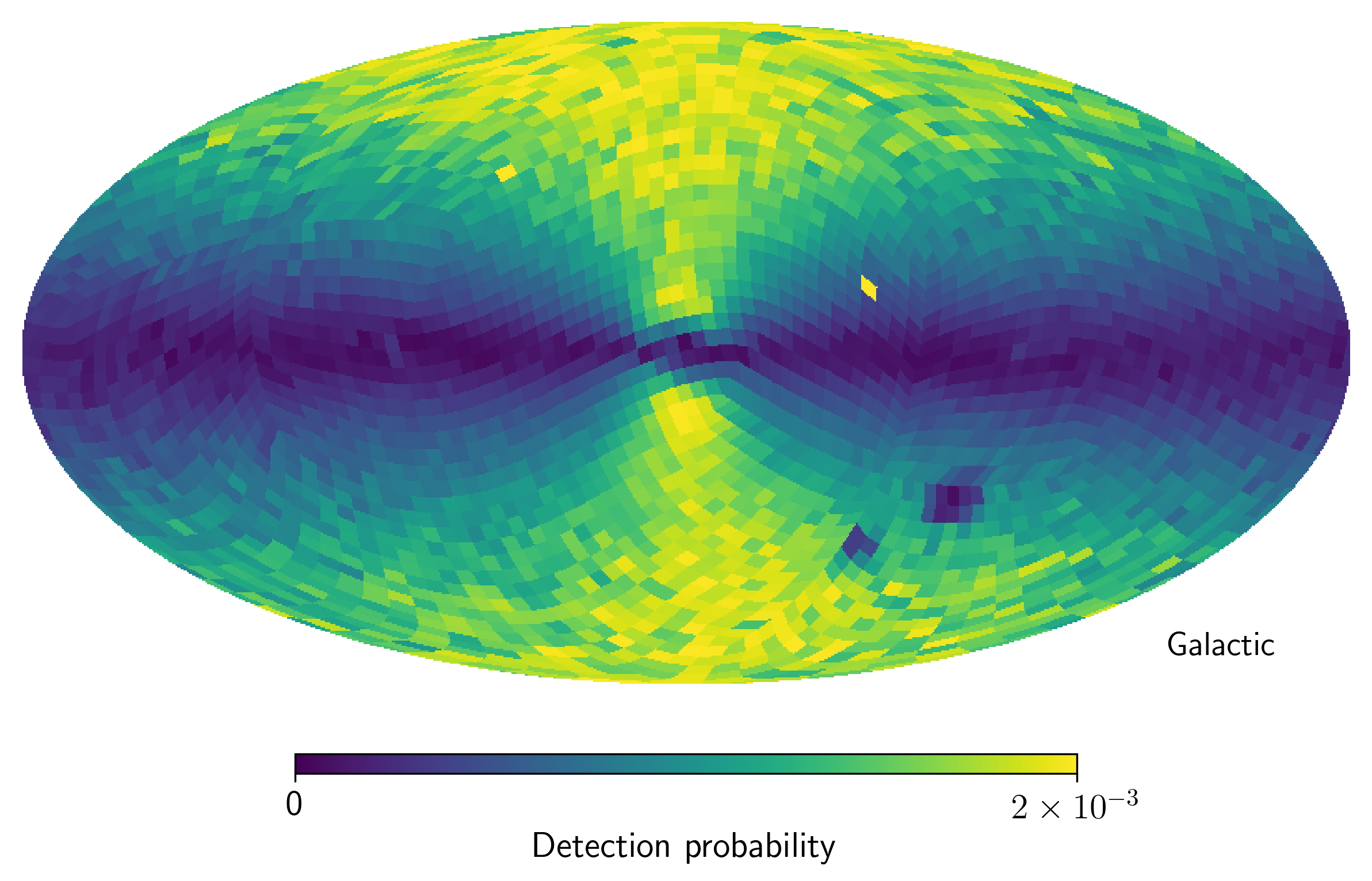}
\caption{Selection function for the sources selected to be part of GS/E, described by the sources with radial velocities, $\varpi > 0.1$ mas, $\varpi/\sigma_\varpi > 5$ mas and $-1500 < \text{L}_z < 150$ kpc km$/$s, with respect to \Gaia DR2 and DR3 for top and bottom panels respectively. The top panel belongs to the original sample in \Gaia DR2 described by \citet{2018Natur.563...85H}, while the bottom panel shows the same cuts applied to \Gaia DR3. The maps correspond to the resolution of HEALPix level $4$. The magnitude and colour dependencies have been marginalised out.}
\label{fig:GES_DR2_DR3}
\end{figure}

In addition to the two parallax cuts whose selection function is displayed in Fig.~\ref{fig:SF_GES}, \citet{2018Natur.563...85H} performed an angular momentum cut retaining only stars with $-1500 < \text{L}_{\text{z}} < 150$ kpc km$/$s. Figure~\ref{fig:GES_DR2_DR3} shows the selection functions for the cuts to produce the GS/E sample relative to \Gaia DR2 and DR3, in the top and bottom panels respectively. Much of the asymmetry seen in the \Gaia DR2 sample is removed in DR3, due to a combination of the better parallax precision and the larger volume explored by the stars with radial velocities. On the other hand, Fig.~\ref{fig:GES_DR2_DR3} shows for \Gaia DR3 a stronger selection effect favouring stars in the Galactic halo. The query to generate the relevant data to compute such a selection function differs from the query in Sect.~\ref{sec:rvs} due to the computation L$_{\text{z}}$, which is done outside the \Gaia archive. The selection function is estimated from a sample of GS/E stars uploaded to the \Gaia archive as a user table, which is crossmatched to the \Gaia source table by \verb+source_id+. The query is provided in App.~\ref{app:appendix}.

\section{Summary}
\label{sec:conclusions}

We have developed a method to estimate the selection function, and its uncertainty, for subsets of the \Gaia data. The methodology provides the means to compute the probability that a source with certain attributes is included in a \textit{sub}-sample provided the \textit{sub}-sample is completely contained in the \Gaia catalogue. To obtain the total selection function this probability should be multiplied with the \Gaia parent catalogue selection function \citep{2023A&A...669A..55C}. Our method is computationally cheap \citep[compared to previous methodologies for the same purpose,][]{2022MNRAS.509.6205E}, which allows for the fast computation of \textit{sub}-sample selection functions from the application of cuts on the \Gaia archive or user-generated data tables (see Appendix~\ref{app:appendix}). The whole method, together with full documentation, is provided in the python package of the GaiaUnlimited project as a customisable class, \texttt{SubsampleSelectionFunction} (see Appendix~\ref{app:github} for a usage example).

We applied the described methodology to estimate the selection function of the subset of \gdr{3} with heliocentric radial velocity measurements, which are also provided as built-in functions in the GaiaUnlimited package. We find that the selection function for the stars with radial velocities is well-constrained for well-populated bins in either the targeted \textit{sub}-sample or the full catalogue (high $k$ and $n$ in our  notation), and less reliable when these numbers are low (as captured by the uncertainty in the selection function, see Fig.~\ref{fig:rvs_var}). For low values of $k$ and $n$, we also characterised the bias of our estimation of the selection probability in Fig.~\ref{fig:bias}, which can also help in selecting the binning of the data in order to have a minimum $n$ in each bin. The main dependencies of the RV selection function are $l$, $b$ and $G$ (following the main dependencies of the \Gaia catalogue selection function) plus the additional dependence of $G-\grp$\footnote{Note, however, that when using the GaiaUnlimited package the user is free to select their own dependencies for the \texttt{SubsampleSelectionFunction} class in the form of an input dictionary.}. The addition of the colour as an argument of the selection function is to capture the temperature and the $G_{RVS}$ dependencies of the RV sample. We assume that the \Gaia catalogue selection function depends on sky position $(l,b)$ and $G$ magnitude. A discussion on the inclusion of a colour dependency in the \Gaia catalogue selection function is out of the scope of this paper. However, as noted by \citet{2023A&A...669A..55C}, no evidence of such dependency was found (see their Sect.~4 for a detailed discussion).

Finally, we estimated the selection function for the different cuts applied by \citet{2018Natur.563...85H} to \gdr{2} data to select the debris of GS/E. When comparing the \Gaia DR2 selection function for the GS/E stars to \Gaia DR3, and as already pointed out by the authors, we find that the asymmetry seen in the structure of GS/E is due to the cuts in parallax ($\varpi > 0.1$ mas) and parallax quality ($\varpi/\sigma_{\varpi} > 5$ mas) rather than an intrinsic structure of the debris. We confirmed this by simulating a spherical distribution of RGB stars and recovering an asymmetric distribution after the application of the selection function. This asymmetry, even though it is still present, is less prominent when the GS/E stars are selected from \gdr{3}.

\begin{acknowledgements}

We thank David W. Hogg for his contributions to the GaiaUnlimited project. This work is a result of the GaiaUnlimited project, which has
received funding from the European Union’s Horizon 2020
research and innovation program under grant agreement No
101004110. The GaiaUnlimited project was started at the 2019
Santa Barbara Gaia Sprint, hosted by the Kavli Institute for Theoretical Physics at the University of California, Santa Barbara.

This work has made use of results from the European Space Agency (ESA)
space mission {\it Gaia}, the data from which were processed by the {\it Gaia
Data Processing and Analysis Consortium} (DPAC).  Funding for the DPAC
has been provided by national institutions, in particular, the
institutions participating in the {\it Gaia} Multilateral Agreement. The
{\it Gaia} mission website is \url{http: //www.cosmos.esa.int/gaia}. The
authors are current or past members of the ESA {\it Gaia} mission team and
of the {\it Gaia} DPAC.

\end{acknowledgements}

\bibliographystyle{aa}
\bibliography{bibliography}

\begin{thebibliography}{18}
\expandafter\ifx\csname natexlab\endcsname\relax\def\natexlab#1{#1}\fi

\bibitem[{{Belokurov} {et~al.}(2018){Belokurov}, {Erkal}, {Evans}, {Koposov},
  \& {Deason}}]{2018MNRAS.478..611B}
{Belokurov}, V., {Erkal}, D., {Evans}, N.~W., {Koposov}, S.~E., \& {Deason},
  A.~J. 2018, \mnras, 478, 611

\bibitem[{{Boubert} \& {Everall}(2020)}]{2020MNRAS.497.4246B}
{Boubert}, D. \& {Everall}, A. 2020, \mnras, 497, 4246

\bibitem[{{Boubert} \& {Everall}(2022)}]{2022MNRAS.510.4626B}
{Boubert}, D. \& {Everall}, A. 2022, \mnras, 510, 4626

\bibitem[{{Boubert} {et~al.}(2021){Boubert}, {Everall}, {Fraser}, {Gration}, \&
  {Holl}}]{2021MNRAS.501.2954B}
{Boubert}, D., {Everall}, A., {Fraser}, J., {Gration}, A., \& {Holl}, B. 2021,
  \mnras, 501, 2954

\bibitem[{{Boubert} {et~al.}(2020){Boubert}, {Everall}, \&
  {Holl}}]{2020MNRAS.497.1826B}
{Boubert}, D., {Everall}, A., \& {Holl}, B. 2020, \mnras, 497, 1826

\bibitem[{{Cantat-Gaudin} {et~al.}(2023){Cantat-Gaudin}, {Fouesneau}, {Rix},
  {Brown}, {Castro-Ginard}, {Kostrzewa-Rutkowska}, {Drimmel}, {Hogg}, {Casey},
  {Khanna}, {Oh}, {Price-Whelan}, {Belokurov}, {Saydjari}, \&
  {Green}}]{2023A&A...669A..55C}
{Cantat-Gaudin}, T., {Fouesneau}, M., {Rix}, H.-W., {et~al.} 2023, \aap, 669,
  A55

\bibitem[{{Everall} \& {Boubert}(2022)}]{2022MNRAS.509.6205E}
{Everall}, A. \& {Boubert}, D. 2022, \mnras, 509, 6205

\bibitem[{{Gaia Collaboration} {et~al.}(2018){Gaia Collaboration}, {Brown},
  {Vallenari}, {Prusti}, {de Bruijne}, {Babusiaux}, {Bailer-Jones}, {Biermann},
  {Evans}, {Eyer}, \& et~al.}]{2018A&A...616A...1G}
{Gaia Collaboration}, {Brown}, A.~G.~A., {Vallenari}, A., {et~al.} 2018, \aap,
  616, A1

\bibitem[{{Gaia Collaboration} {et~al.}(2016){Gaia Collaboration}, {Prusti},
  {de Bruijne}, {Brown}, {Vallenari}, {Babusiaux}, {Bailer-Jones}, {Bastian},
  {Biermann}, {Evans}, \& et~al.}]{2016A&A...595A...1G}
{Gaia Collaboration}, {Prusti}, T., {de Bruijne}, J.~H.~J., {et~al.} 2016,
  \aap, 595, A1

\bibitem[{{Gaia Collaboration} {et~al.}(2022){Gaia Collaboration}, {Vallenari},
  {Brown}, {Prusti}, {de Bruijne}, {Arenou}, {Babusiaux}, {Biermann},
  {Creevey}, {Ducourant}, {Evans}, {Eyer}, {Guerra}, {Hutton}, {Jordi},
  {Klioner}, {Lammers}, {Lindegren}, {Luri}, {Mignard}, {Panem}, {Pourbaix},
  {Randich}, {Sartoretti}, {Soubiran}, {Tanga}, {Walton}, {Bailer-Jones},
  {Bastian}, {Drimmel}, {Jansen}, {Katz}, {Lattanzi}, {van Leeuwen}, {Bakker},
  {Cacciari}, {Casta{\~n}eda}, {De Angeli}, {Fabricius}, {Fouesneau},
  {Fr{\'e}mat}, {Galluccio}, {Guerrier}, {Heiter}, {Masana}, {Messineo},
  {Mowlavi}, {Nicolas}, {Nienartowicz}, {Pailler}, {Panuzzo}, {Riclet}, {Roux},
  {Seabroke}, {Sordo{\o}rcit}, {Th{\'e}venin}, {Gracia-Abril}, {Portell},
  {Teyssier}, {Altmann}, {Andrae}, {Audard}, {Bellas-Velidis}, {Benson},
  {Berthier}, {Blomme}, {Burgess}, {Busonero}, {Busso}, {C{\'a}novas}, {Carry},
  {Cellino}, {Cheek}, {Clementini}, {Damerdji}, {Davidson}, {de Teodoro},
  {Nu{\~n}ez Campos}, {Delchambre}, {Dell'Oro}, {Esquej},
  {Fern{\'a}ndez-Hern{\'a}ndez}, {Fraile}, {Garabato}, {Garc{\'\i}a-Lario},
  {Gosset}, {Haigron}, {Halbwachs}, {Hambly}, {Harrison}, {Hern{\'a}ndez},
  {Hestroffer}, {Hodgkin}, {Holl}, {Jan{\ss}en}, {Jevardat de Fombelle},
  {Jordan}, {Krone-Martins}, {Lanzafame}, {L{\"o}ffler}, {Marchal}, {Marrese},
  {Moitinho}, {Muinonen}, {Osborne}, {Pancino}, {Pauwels}, {Recio-Blanco},
  {Reyl{\'e}}, {Riello}, {Rimoldini}, {Roegiers}, {Rybizki}, {Sarro}, {Siopis},
  {Smith}, {Sozzetti}, {Utrilla}, {van Leeuwen}, {Abbas}, {{\'A}brah{\'a}m},
  {Abreu Aramburu}, {Aerts}, {Aguado}, {Ajaj}, {Aldea-Montero}, {Altavilla},
  {{\'A}lvarez}, {Alves}, {Anders}, {Anderson}, {Anglada Varela}, {Antoja},
  {Baines}, {Baker}, {Balaguer-N{\'u}{\~n}ez}, {Balbinot}, {Balog}, {Barache},
  {Barbato}, {Barros}, {Barstow}, {Bartolom{\'e}}, {Bassilana}, {Bauchet},
  {Becciani}, {Bellazzini}, {Berihuete}, {Bernet}, {Bertone}, {Bianchi},
  {Binnenfeld}, {Blanco-Cuaresma}, {Blazere}, {Boch}, {Bombrun}, {Bossini},
  {Bouquillon}, {Bragaglia}, {Bramante}, {Breedt}, {Bressan}, {Brouillet},
  {Brugaletta}, {Bucciarelli}, {Burlacu}, {Butkevich}, {Buzzi}, {Caffau},
  {Cancelliere}, {Cantat-Gaudin}, {Carballo}, {Carlucci}, {Carnerero},
  {Carrasco}, {Casamiquela}, {Castellani}, {Castro-Ginard}, {Chaoul},
  {Charlot}, {Chemin}, {Chiaramida}, {Chiavassa}, {Chornay}, {Comoretto},
  {Contursi}, {Cooper}, {Cornez}, {Cowell}, {Crifo}, {Cropper}, {Crosta},
  {Crowley}, {Dafonte}, {Dapergolas}, {David}, {David}, {de Laverny}, {De
  Luise}, {De March}, {De Ridder}, {de Souza}, {de Torres}, {del Peloso}, {del
  Pozo}, {Delbo}, {Delgado}, {Delisle}, {Demouchy}, {Dharmawardena}, {Di
  Matteo}, {Diakite}, {Diener}, {Distefano}, {Dolding}, {Edvardsson}, {Enke},
  {Fabre}, {Fabrizio}, {Faigler}, {Fedorets}, {Fernique}, {Fienga}, {Figueras},
  {Fournier}, {Fouron}, {Fragkoudi}, {Gai}, {Garcia-Gutierrez},
  {Garcia-Reinaldos}, {Garc{\'\i}a-Torres}, {Garofalo}, {Gavel}, {Gavras},
  {Gerlach}, {Geyer}, {Giacobbe}, {Gilmore}, {Girona}, {Giuffrida}, {Gomel},
  {Gomez}, {Gonz{\'a}lez-N{\'u}{\~n}ez}, {Gonz{\'a}lez-Santamar{\'\i}a},
  {Gonz{\'a}lez-Vidal}, {Granvik}, {Guillout}, {Guiraud},
  {Guti{\'e}rrez-S{\'a}nchez}, {Guy}, {Hatzidimitriou}, {Hauser}, {Haywood},
  {Helmer}, {Helmi}, {Sarmiento}, {Hidalgo}, {Hilger}, {H{\l}adczuk}, {Hobbs},
  {Holland}, {Huckle}, {Jardine}, {Jasniewicz}, {Jean-Antoine Piccolo},
  {Jim{\'e}nez-Arranz}, {Jorissen}, {Juaristi Campillo}, {Julbe}, {Karbevska},
  {Kervella}, {Khanna}, {Kontizas}, {Kordopatis}, {Korn}, {K{\'o}sp{\'a}l},
  {Kostrzewa-Rutkowska}, {Kruszy{\'n}ska}, {Kun}, {Laizeau}, {Lambert},
  {Lanza}, {Lasne}, {Le Campion}, {Lebreton}, {Lebzelter}, {Leccia}, {Leclerc},
  {Lecoeur-Taibi}, {Liao}, {Licata}, {Lindstr{\o}m}, {Lister}, {Livanou},
  {Lobel}, {Lorca}, {Loup}, {Madrero Pardo}, {Magdaleno Romeo}, {Managau},
  {Mann}, {Manteiga}, {Marchant}, {Marconi}, {Marcos}, {Marcos Santos},
  {Mar{\'\i}n Pina}, {Marinoni}, {Marocco}, {Marshall}, {Polo},
  {Mart{\'\i}n-Fleitas}, {Marton}, {Mary}, {Masip}, {Massari},
  {Mastrobuono-Battisti}, {Mazeh}, {McMillan}, {Messina}, {Michalik}, {Millar},
  {Mints}, {Molina}, {Molinaro}, {Moln{\'a}r}, {Monari}, {Mongui{\'o}},
  {Montegriffo}, {Montero}, {Mor}, {Mora}, {Morbidelli}, {Morel}, {Morris},
  {Muraveva}, {Murphy}, {Musella}, {Nagy}, {Noval}, {Oca{\~n}a}, {Ogden},
  {Ordenovic}, {Osinde}, {Pagani}, {Pagano}, {Palaversa}, {Palicio},
  {Pallas-Quintela}, {Panahi}, {Payne-Wardenaar}, {Pe{\~n}alosa Esteller},
  {Penttil{\"a}}, {Pichon}, {Piersimoni}, {Pineau}, {Plachy}, {Plum}, {Poggio},
  {Pr{\v{s}}a}, {Pulone}, {Racero}, {Ragaini}, {Rainer}, {Raiteri}, {Rambaux},
  {Ramos}, {Ramos-Lerate}, {Re Fiorentin}, {Regibo}, {Richards}, {Rios Diaz},
  {Ripepi}, {Riva}, {Rix}, {Rixon}, {Robichon}, {Robin}, {Robin}, {Roelens},
  {Rogues}, {Rohrbasser}, {Romero-G{\'o}mez}, {Rowell}, {Royer}, {Ruz Mieres},
  {Rybicki}, {Sadowski}, {S{\'a}ez N{\'u}{\~n}ez}, {Sagrist{\`a} Sell{\'e}s},
  {Sahlmann}, {Salguero}, {Samaras}, {Sanchez Gimenez}, {Sanna},
  {Santove{\~n}a}, {Sarasso}, {Schultheis}, {Sciacca}, {Segol}, {Segovia},
  {S{\'e}gransan}, {Semeux}, {Shahaf}, {Siddiqui}, {Siebert}, {Siltala},
  {Silvelo}, {Slezak}, {Slezak}, {Smart}, {Snaith}, {Solano}, {Solitro},
  {Souami}, {Souchay}, {Spagna}, {Spina}, {Spoto}, {Steele},
  {Steidelm{\"u}ller}, {Stephenson}, {S{\"u}veges}, {Surdej}, {Szabados},
  {Szegedi-Elek}, {Taris}, {Taylo}, {Teixeira}, {Tolomei}, {Tonello}, {Torra},
  {Torra}, {Torralba Elipe}, {Trabucchi}, {Tsounis}, {Turon}, {Ulla}, {Unger},
  {Vaillant}, {van Dillen}, {van Reeven}, {Vanel}, {Vecchiato}, {Viala},
  {Vicente}, {Voutsinas}, {Weiler}, {Wevers}, {Wyrzykowski}, {Yoldas}, {Yvard},
  {Zhao}, {Zorec}, {Zucker}, \& {Zwitter}}]{2022arXiv220800211G}
{Gaia Collaboration}, {Vallenari}, A., {Brown}, A.~G.~A., {et~al.} 2022, arXiv
  e-prints, arXiv:2208.00211

\bibitem[{{Helmi} {et~al.}(2018){Helmi}, {Babusiaux}, {Koppelman}, {Massari},
  {Veljanoski}, \& {Brown}}]{2018Natur.563...85H}
{Helmi}, A., {Babusiaux}, C., {Koppelman}, H.~H., {et~al.} 2018, \nat, 563, 85

\bibitem[{{Katz} {et~al.}(2022){Katz}, {Sartoretti}, {Guerrier}, {Panuzzo},
  {Seabroke}, {Th{\'e}venin}, {Cropper}, {Benson}, {Blomme}, {Haigron},
  {Marchal}, {Smith}, {Baker}, {Chemin}, {Damerdji}, {David}, {Dolding},
  {Fr{\'e}mat}, {Gosset}, {Jan{\ss}en}, {Jasniewicz}, {Lobel}, {Plum},
  {Samaras}, {Snaith}, {Soubiran}, {Vanel}, {Zwitter}, {Antoja}, {Arenou},
  {Babusiaux}, {Brouillet}, {Caffau}, {Di Matteo}, {Fabre}, {Fabricius},
  {Frakgoudi}, {Haywood}, {Huckle}, {Hottier}, {Lasne}, {Leclerc},
  {Mastrobuono-Battisti}, {Royer}, {Teyssier}, {Zorec}, {Crifo}, {Jean-Antoine
  Piccolo}, {Turon}, \& {Viala}}]{2022arXiv220605902K}
{Katz}, D., {Sartoretti}, P., {Guerrier}, A., {et~al.} 2022, arXiv e-prints,
  arXiv:2206.05902

\bibitem[{{Riello} {et~al.}(2021){Riello}, {De Angeli}, {Evans}, {Montegriffo},
  {Carrasco}, {Busso}, {Palaversa}, {Burgess}, {Diener}, {Davidson}, {Rowell},
  {Fabricius}, {Jordi}, {Bellazzini}, {Pancino}, {Harrison}, {Cacciari}, {van
  Leeuwen}, {Hambly}, {Hodgkin}, {Osborne}, {Altavilla}, {Barstow}, {Brown},
  {Castellani}, {Cowell}, {De Luise}, {Gilmore}, {Giuffrida}, {Hidalgo},
  {Holland}, {Marinoni}, {Pagani}, {Piersimoni}, {Pulone}, {Ragaini}, {Rainer},
  {Richards}, {Sanna}, {Walton}, {Weiler}, \& {Yoldas}}]{2021A&A...649A...3R}
{Riello}, M., {De Angeli}, F., {Evans}, D.~W., {et~al.} 2021, \aap, 649, A3

\bibitem[{{Rix} {et~al.}(2021){Rix}, {Hogg}, {Boubert}, {Brown}, {Casey},
  {Drimmel}, {Everall}, {Fouesneau}, \& {Price-Whelan}}]{2021AJ....162..142R}
{Rix}, H.-W., {Hogg}, D.~W., {Boubert}, D., {et~al.} 2021, \aj, 162, 142

\bibitem[{{Rybizki} {et~al.}(2021){Rybizki}, {Rix}, {Demleitner},
  {Bailer-Jones}, \& {Cooper}}]{2021MNRAS.500..397R}
{Rybizki}, J., {Rix}, H.-W., {Demleitner}, M., {Bailer-Jones}, C. A.~L., \&
  {Cooper}, W.~J. 2021, \mnras, 500, 397

\bibitem[{{Sartoretti} {et~al.}(2022){Sartoretti}, {Marchal}, {Babusiaux},
  {Jordi}, {Guerrier}, {Panuzzo}, {Katz}, {Seabroke}, {Th{\'e}venin},
  {Cropper}, {Benson}, {Blomme}, {Haigron}, {Smith}, {Baker}, {Chemin},
  {David}, {Dolding}, {Fr{\'e}mat}, {Janssen}, {Jasniewicz}, {Lobel}, {Plum},
  {Samaras}, {Snaith}, {Soubiran}, {Vanel}, {Zwitter}, {Brouillet}, {Caffau},
  {Crifo}, {Fabre}, {Frakgoudi}, {Jean-Antoine Piccolo}, {Huckle}, {Lasne},
  {Leclerc}, {Mastrobuono-Battisti}, {Royer}, {Viala}, \&
  {Zorec}}]{2022arXiv220605725S}
{Sartoretti}, P., {Marchal}, O., {Babusiaux}, C., {et~al.} 2022, arXiv
  e-prints, arXiv:2206.05725

\bibitem[{{Saydjari} {et~al.}(2023){Saydjari}, {Schlafly}, {Lang}, {Meisner},
  {Green}, {Zucker}, {Zelko}, {Speagle}, {Daylan}, {Lee}, {Valdes}, {Schlegel},
  \& {Finkbeiner}}]{2023ApJS..264...28S}
{Saydjari}, A.~K., {Schlafly}, E.~F., {Lang}, D., {et~al.} 2023, \apjs, 264, 28

\bibitem[{{Schlafly} {et~al.}(2018){Schlafly}, {Green}, {Lang}, {Daylan},
  {Finkbeiner}, {Lee}, {Meisner}, {Schlegel}, \& {Valdes}}]{Schlafly18}
{Schlafly}, E.~F., {Green}, G.~M., {Lang}, D., {et~al.} 2018, \apjs, 234, 39

\end{thebibliography}

\onecolumn
\begin{appendix}
\section{Example query to the \Gaia archive}
\label{app:appendix}

The query below is an example of how to retrieve a table from the \Gaia archive, with the needed information binned in sky position (HEALPix level $5$, from \verb+source_id+), magnitude $G$ and colour $G-\grp$. In this particular case, we show how to query the stars with available \verb+radial_velocity+  measurements. The $G$ magnitude is binned from $3$ to $20$ in bins of $0.2$ mag. For the $G-\grp$ colour, the bin size is $0.4$ mag in the range of $-2.5$ to $5.1$.

The result of the query is shown in Table~\ref{tab:resultquery} where, for a particular HEALPix pixel and $G$ and $G-\grp$ bin numbers, the total number of stars ($n$) and the number of stars fulfilling the specific selection ($k$) is retrieved.

\begin{verbatim}
SELECT magnitude, colour, position, COUNT(*) AS n, SUM(selection) AS k
FROM (SELECT to_integer(floor((phot_g_mean_mag - 3)/0.2)) AS magnitude,
             to_integer(floor((g_rp + 2.5)/0.4)) AS colour, 
             to_integer(GAIA_HEALPIX_INDEX(5, source_id)) AS position,
             to_integer(IF_THEN_ELSE('radial_velocity is not null', 1.0,0.0)) AS selection
      FROM gaiadr3.gaia_source 
      WHERE phot_g_mean_mag > 3 AND phot_g_mean_mag < 20
      AND g_rp > -2.5 AND g_rp < 5.1) AS subquery
GROUP BY magnitude, colour, position
\end{verbatim}

\begin{table}[hb!]
\begin{center}
	\caption{ \label{tab:resultquery} Values for the total number of stars in \Gaia ($n$) and the number of stars with heliocentric radial velocity measurements ($k$) binned in sky position (HEALPix index at level $5$), magnitude $G$ and colour $G-\grp$.}

	\begin{tabular}{ c  c  c  c  c }
	\hline
	\hline

    HEALPix index & Magnitude bin &  Colour bin & $n$ & $k$\\
        \hline
$0$ & $25$ & $7$ & $1$ & $1$ \\
$0$ & $34$ & $6$ & $1$ & $1$ \\
$0$ & $34$ & $7$ & $2$ & $1$ \\
\\
\multicolumn{5}{c}{$\vdots$} \\
\\
$196607$ & $84$ & $10$ & $1$ & $0$ \\
$196607$ & $84$ & $12$ & $1$ & $0$ \\
	\hline
	\hline
	\end{tabular}
\end{center}
\end{table}

Alternatively, if the selection function to be computed is that from a user-made table, given that all the sources are in the \Gaia catalogue, the query to be performed in the \Gaia archive relies on a crossmatch on \verb+source_id+. The following query provides an example based on Sect.~\ref{sec:GES}, where the L$_z$ has been computed outside the \Gaia archive and the resulting sample has been uploaded as \verb+user_acastr01.ges_dr3+.

\begin{verbatim}
SELECT magnitude, colour, position, COUNT(*) AS n, SUM(selection) AS k
FROM (SELECT to_integer(floor((phot_g_mean_mag - 3)/0.2)) AS magnitude,
             to_integer(floor((g_rp + 2.5)/0.4)) AS colour, 
             to_integer(GAIA_HEALPIX_INDEX(4, source_id)) AS position,
             to_integer(IF_THEN_ELSE(
                        'source_id in (select source_id from user_acastr01.ges_dr3)', 1.0,0.0)
      ) AS selection
      FROM gaiadr2.gaia_source 
      WHERE phot_g_mean_mag > 3 AND phot_g_mean_mag < 20
      AND g_rp > -2.5 AND g_rp < 5.1) AS subquery
GROUP BY magnitude, colour, position
\end{verbatim}

\clearpage

\section{Comparison of the \Gaia EDR3 and DR3 RV selection function}
\label{app:dr2_vs_dr3}

In this section, we compare the completeness of the RV sample in \Gaia EDR3 with respect to \Gaia DR3, at magnitude $G = 13$ which is beyond the \Gaia EDR3 magnitude limit. The coverage of the RV sample in \Gaia DR3 has greatly improved, and the features such as the IGSL seen in \Gaia EDR3 are removed.

\begin{figure*}[ht!]
\centering
\includegraphics[width = 0.85\textwidth]{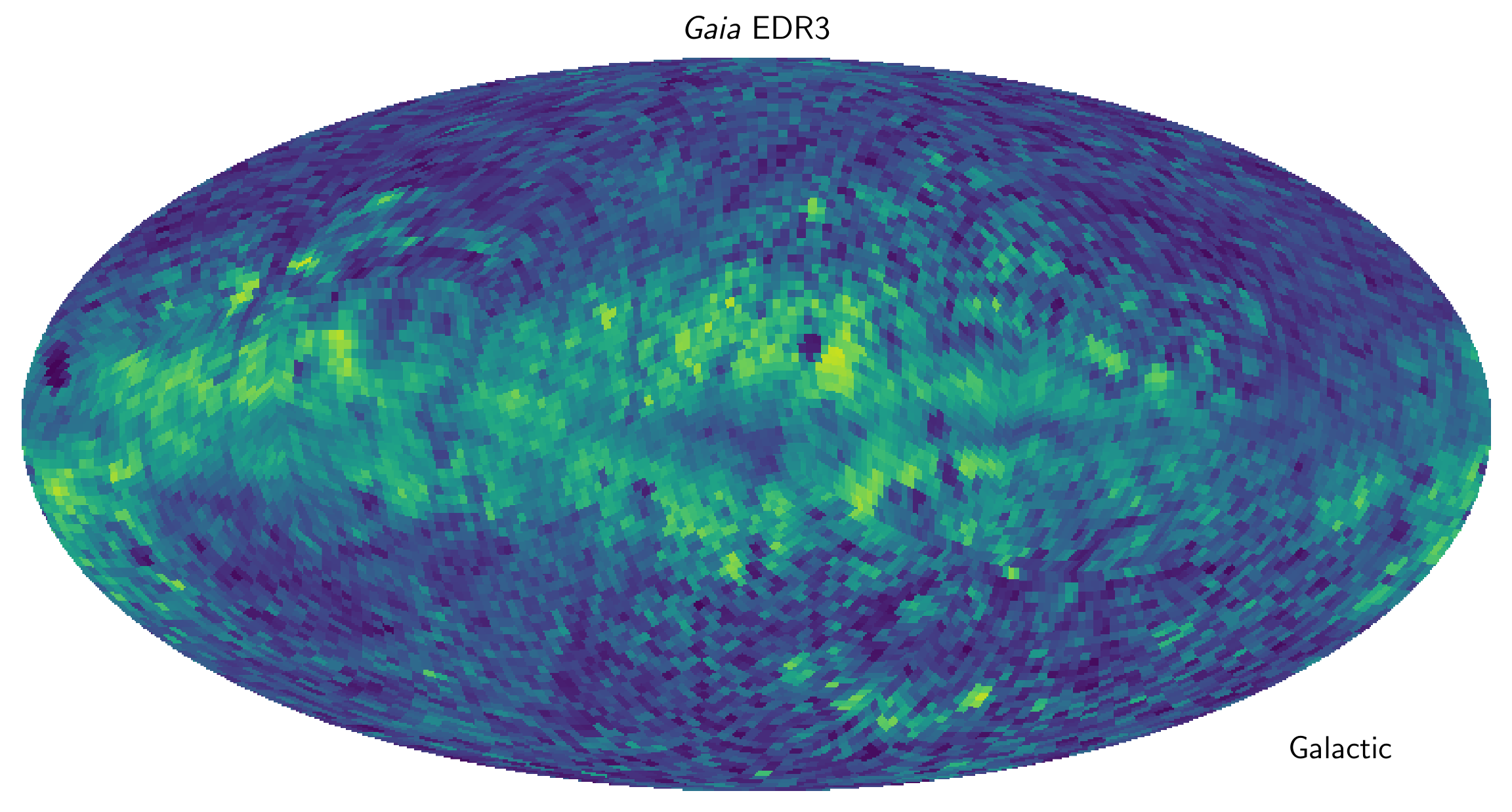}
\includegraphics[width = 0.85\textwidth]{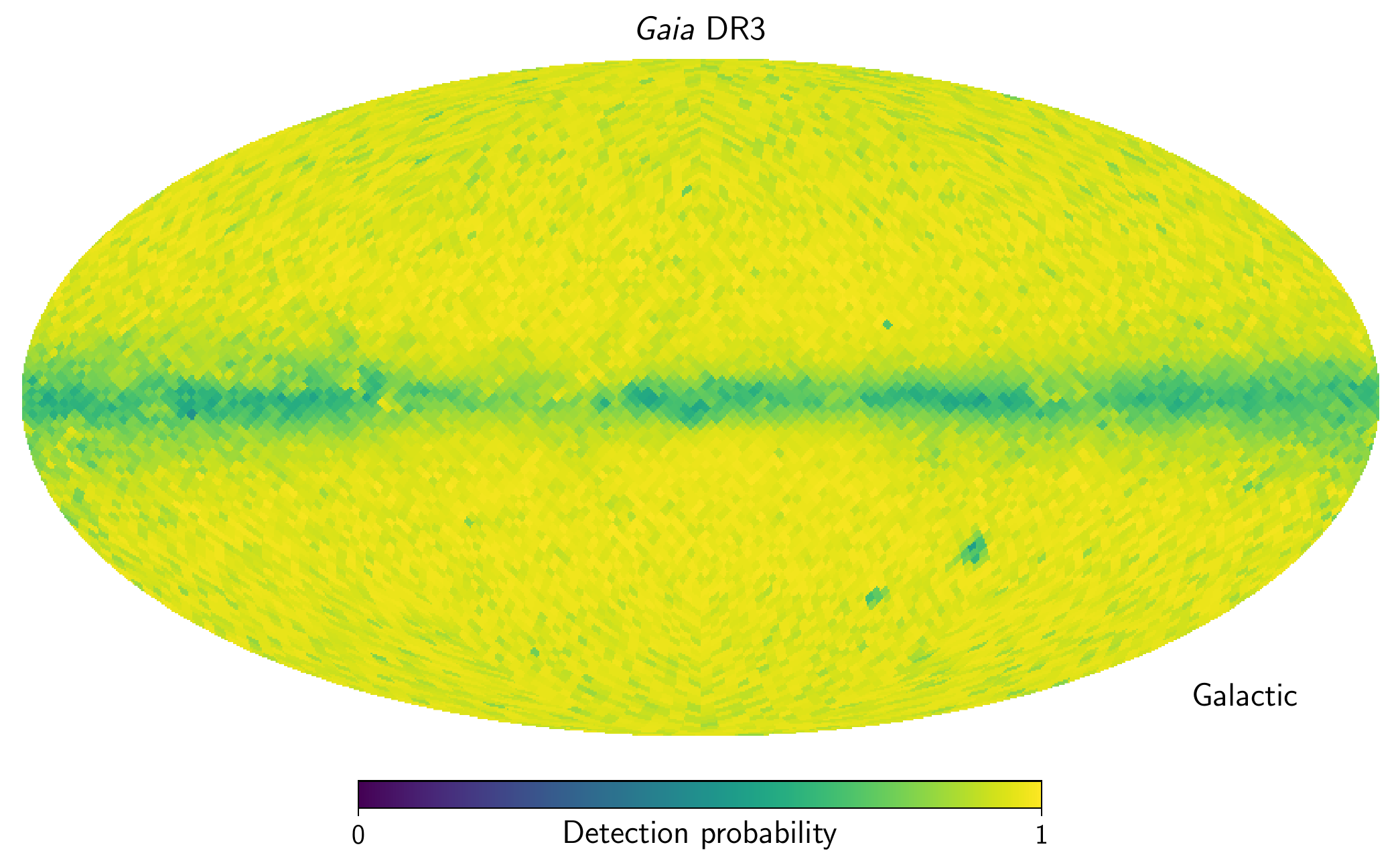}
\caption{Completeness sky map at HEALPix level $5$ of the sources with available radial velocities at magnitude $G = 13$ in \Gaia EDR3 (top panel) and in \Gaia DR3 (bottom panel). The dependency on colour has been marginalised out.}
\label{fig:dr2_vs_dr3}
\end{figure*}

\clearpage

\section{Using the GaiaUnlimited package}
\label{app:github}

As part of the GaiaUnlimited python package, we provide the class \texttt{SubsampleSelectionFunction} to generate selection functions for \textit{sub}-samples from the \Gaia catalogue. The user is expected to  provide the \Gaia archive query that produces the \textit{sub}-sample. Then, the selection function for the \textit{sub}-sample is calculated according to the methodology outlined in Sect.~\ref{sec:method}. 

The \texttt{SubsampleSelectionFunction} class takes three arguments: \texttt{subsample\_query}, which is the query to be performed in the \Gaia archive; \texttt{file\_name}, which is used to store the data resulting from the query and save time in following executions of the same run; and a python dictionary including the dimensions in which to bin the data and their binning. This dictionary should include the desired HEALPix level, and additional \Gaia columns to bin the data. For instance, to generate a selection function at the resolution of HEALPix level $5$, $G \in [3,20]$ in steps of $0.2$ and $G-\grp \in [-2.5,5.1]$ in bins of $0.4$, the expected dictionary is: 
\begin{verbatim}
    inDict = {'healpix': 5, 'phot_g_mean_mag': [3,20,0.2], 'g_rp': [-2.5,5.1,0.4]}.
\end{verbatim}
Additional columns as additional dependencies in the selection function may be added, bearing in mind that this may increase the execution time (of the query) and decrease the number of stars (in both $k$ and $n$ in our notation) in each bin, therefore increasing the noise and the number of regions of the sky with no available data (see Sect.~\ref{sec:comparison}). 

Once the \texttt{SubsampleSelectionFunction} class has been initialised, the resulting selection function can be queried by providing the targeted coordinates (can be an array with the centres of the HEALPix pixels for an all-sky plot), magnitude $G$, colour $G-\grp$ and the possible additional columns. To access the desired magnitude and colour bins (and any other dimension from \Gaia columns), the name of the column plus an underscore (\_) should be provided as the argument name (an example is shown in Listing~\ref{lst:label}). 

Listing~\ref{lst:label} shows the python code used to generate one of the completeness maps shown in Fig.~\ref{fig:rvs}. The time to execute Listing~\ref{lst:label} is dominated by the query to the archive (line 14) and in this case, is about $40$ min.

We show further applications of the \texttt{SubsampleSelectionFunction} class by estimating the selection function for sources with i) a measured parallax and proper motion and ii) \texttt{RUWE}$< 1.4$. The completeness maps and the relevant change on the main code shown in Listing~\ref{lst:label} are shown in the Appendix~\ref{app:examples}.

\begin{flushright}
\begin{lstlisting}[language=Python, caption=Python code to generate Fig.~\ref{fig:rvs},label={lst:label}]
import healpy as hp
from astroquery.gaia import Gaia
from gaiaunlimited.utils import get_healpix_centers
from gaiaunlimited.selectionfunctions.subsample import SubsampleSelectionFunction

#Login to the Gaia archive to save the query
Gaia.MAIN_GAIA_TABLE = "gaiadr3.gaia_source"
Gaia.login(user = username, password = passwd)

#Define the dependencies and resolutions of the selection function
inDict = {'healpix': 5, 'phot_g_mean_mag': [3,20,0.2], 'g_rp': [-2.5,5.1,0.4]}

#Initiate the SubsampleSelectionFunction class
dr3SubsampleSF = SubsampleSelectionFunction(subsample_query = "radial_velocity is not null", file_name = "radial_velocity", hplevel_and_binning = inDict)

#Select where we want the selection function to be evaluated
healpix_level = 5
G = 13
G_RP = 0.5
coords_of_centers = get_healpix_centers(healpix_level)
gmag = np.ones_like(coords_of_centers) * G
col = np.ones_like(coords_of_centers) * G_RP

#Query the completeness of the subsample
completeness,variance = dr3SubsampleSF.query(coords_of_centers, phot_g_mean_mag_ = gmag, g_rp_ = col,return_variance = True,fill_nan = False)

#Plot the completeness map
hp.mollview(completeness,coord =["Celestial","Galactic"], min=0, max=1, title=f"RV completeness at G = {G:.1f} and G_RP = {G_RP:.1f}")
\end{lstlisting}
\end{flushright}

\clearpage

\section{Example of other selection functions}
\label{app:examples}

\begin{figure*}[htb!]
\centering
\includegraphics[width = 0.67\textwidth]{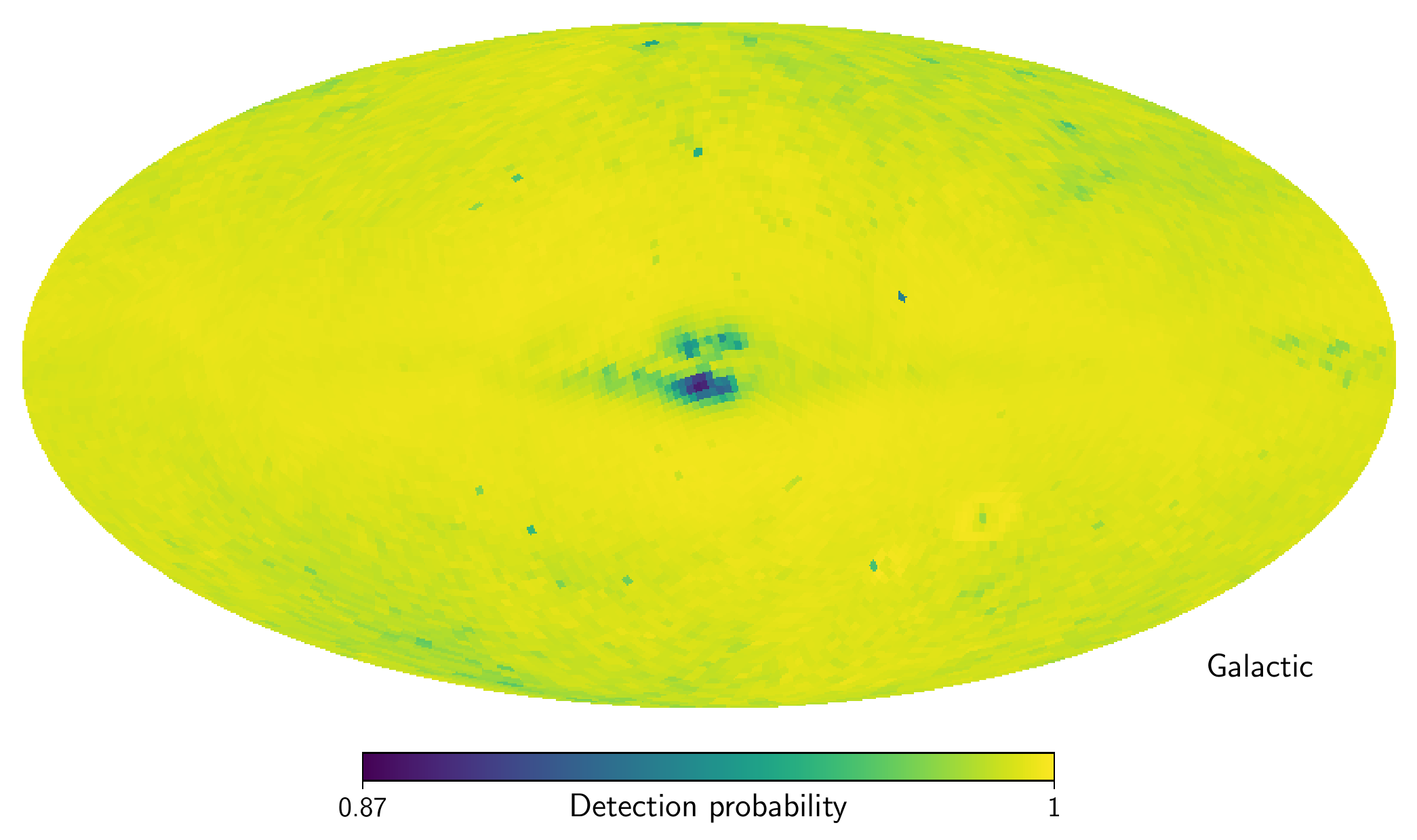}
\caption{Completeness map at HEALPix level $5$ of the sources with reported parallax and proper motions. The magnitude and colour dependencies have been marginalised out.}
\label{fig:par_pm}
\end{figure*}

\begin{figure*}[htb!]
\centering
\includegraphics[width = 0.67\textwidth]{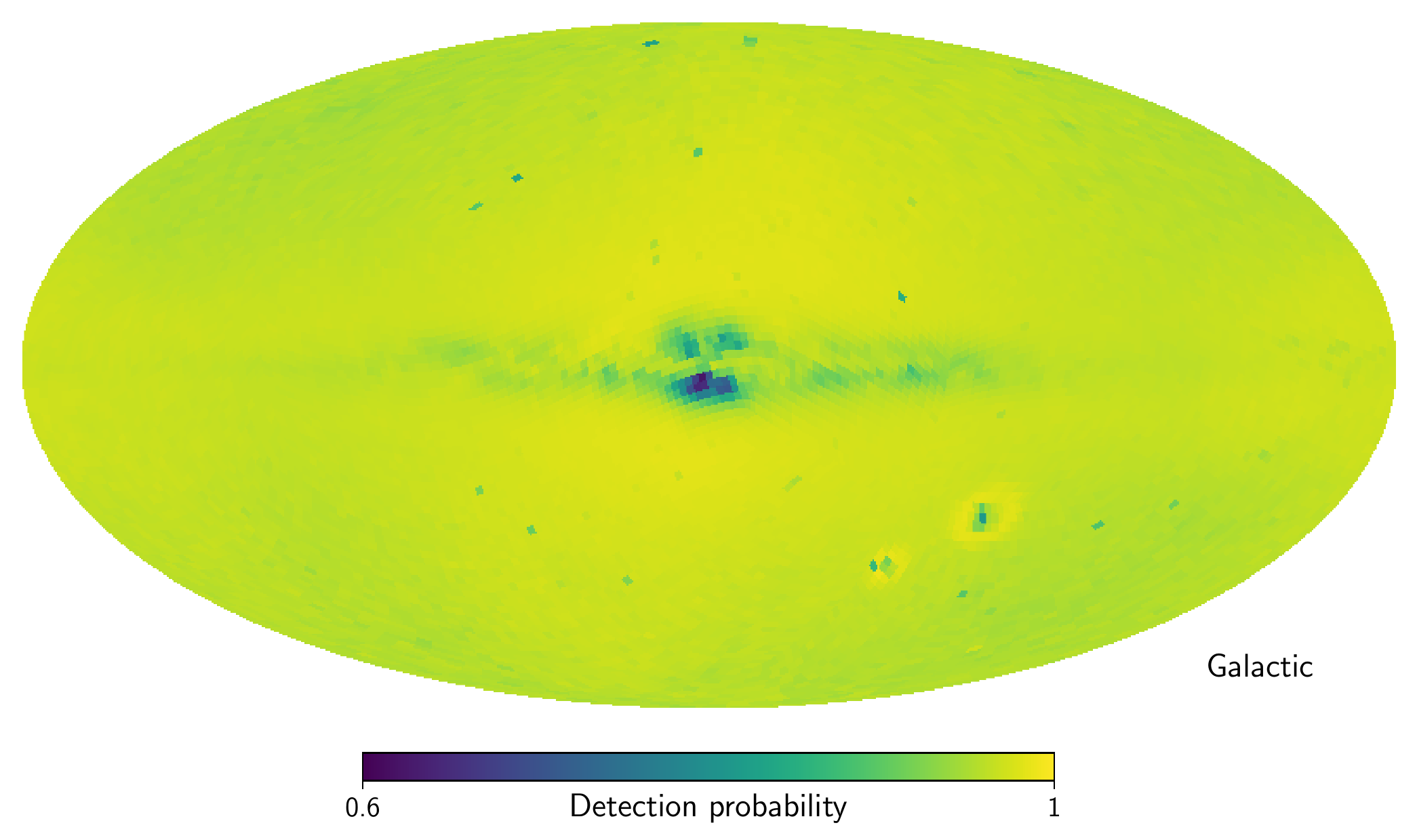}
\caption{Completeness map at HEALPix level $5$ for the sources with \texttt{RUWE} $< 1.4$. The magnitude and colour dependencies have been marginalised out.}
\label{fig:ruwe}
\end{figure*}

Similarly to \citet{2022MNRAS.509.6205E}, in this Appendix we show the completeness maps for the sources with i) parallax and proper motion measurements (Fig.~\ref{fig:par_pm}), and ii) sources with \verb+RUWE+ $< 1.4$ (Fig.~\ref{fig:ruwe}). These selection functions are as a function of sky position only, the dependencies with magnitude and colour have been marginalised out. The code to generate these selection functions and completeness maps is similar to that in Listing~\ref{lst:label}, where the \verb+SubsampleSelectionFunction+ class has been initialised as shown in Listing~\ref{lst:app}.

\begin{flushright}
\begin{minipage}[c]{0.97\columnwidth}
\begin{lstlisting}[language=Python, caption=Initialisation of the \texttt{SubsampleSelectionFunction} class for Fig.~\ref{fig:par_pm} and Fig.~\ref{fig:ruwe},label={lst:app}]
#Sources with reported parallax and proper motions
dr3AstrometrySF = SubsampleSelectionFunction(subsample_query = "parallax is not null and pmra is not null and pmdec is not null",file_name = "par_pm", hplevel_and_binning = inDict)

#Sources with ruwe < 1.4
dr3RUWESF = SubsampleSelectionFunction(subsample_query = "ruwe < 1.4",file_name = "ruwe_1.4",hplevel_and_binning = inDict)
\end{lstlisting}
\end{minipage}
\end{flushright}

\end{appendix}
\end{document}